# Comet 2P/Encke in apparitions of 2013 and 2017: I. Imaging photometry and long-slit spectroscopy

Vera Rosenbush [a,b,*], Oleksandra Ivanova [a,b,c], Valerii Kleshchonok [a], Nikolai Kiselev [b,d], Viktor Afanasiev [e], Olena Shubina [b], Dmitry Petrov [d]

[a] Taras Shevchenko National University of Kyiv, Astronomical Observatory, 3 Observatorna Str., 04053 Kyiv, Ukraine
[b] Main Astronomical Observatory of the National Academy of Sciences of Ukraine, 27 Zabolotnoho Str., 03143 Kyiv, Ukraine
[c] Astronomical Institute of the Slovak Academy of Sciences, SK-05960 Tatranská Lomnica, Slovak Republic
[d] Crimean Astrophysical Observatory, Nauchnij, Crimea, Ukraine
[e] Special Astrophysical Observatory of the Russian Academy of Sciences, 369167 Nizhnij Arkhyz, Russia

A B S T R A C T

We present the results of imaging photometric and long-slit spectroscopic observations of comet 2P/Encke performed at the heliocentric distance $r = 0.56$ au, geocentric distance $\Delta = 0.65$ au, and phase angle $\alpha = 109.2°$ on November 4, 2013 and at $r = 1.05$ au, $\Delta = 1.34$ au, and $\alpha = 46.8°$ on January 23, 2017. Observations were carried out at the 6-m BTA telescope of the Special Astrophysical Observatory (Russia) with the multimode focal reducer SCORPIO-2. In 2013, the direct images of comet Encke were obtained with the broad-band V filters, whereas in 2017 the narrow-band cometary BC, RC, and NH$_2$ filters as well as the medium-band SED500 and broad-band r-sdss filters were used for observations. About 60 emissions belonging to the CN, C$_2$, C$_3$, NH$_2$, CH, and CO$^+$ molecules were identified within the range $\lambda$3750–7100 Å. The ratios of the production rates C$_2$/CN and C$_3$/CN correspond to the typical comets, not depleted in the carbon-chain. A complex structure of the coma was detected in both observational periods. In January 2017, the dust was in general concentrated near the nucleus, the dust/gas ratio was 2.9 in the r-sdss filter, however, this ratio was larger than 1 at distances 3000–40,000 km from the nucleus. We found that about 75% of the flux of the reflected light in the central pixel was due to the nucleus, whereas the nucleus's flux contributed 48% in the total intensity of the 2000 km area of the coma. We found that after correction for the dust coma contamination the nucleus magnitude is $18.8^m \pm 0.2^m$. Corrected for emissions penetrating into the BC and RC filters and nucleus contribution, color index BC–RC decreased sharply from about $1.43^m$ in the innermost near-nucleus coma to $\sim 0.4^m$ at the distance $\sim 2500$ km. Color BC–RC of the cometary nucleus was about $1.39^m$. We performed a dynamical simulation of dust particles to characterize the morphology and found that visible jets in both observational periods were formed by a single active source located in the north hemisphere at the cometocentric latitude $+55°$. The behavior of the dust coma near the nucleus indicates a strong dust transformation at the distances below $\sim 2000$ km.

## Introduction

Based on the Tisserand parameter ($T = 3.03$) and a surprisingly small aphelion distance (4.1 au, i.e. less than that of Jupiter), Levison (1996), and later Levison and Duncan (1997), separated comet 2P/Encke (hereafter Encke) from the Jupiter-family comets to a specified class comets called the Encke-type. This comet has the shortest orbital period of all known comets, equal to 3.28 years, and in 2017 was observed at its 70th apparition and this is the highest number of any comet. Due to its unique orbital properties, Encke is one of the best-studied short-period comets in photometric and spectral sense. Photometric and spectral observations of the comet in the visible and infrared domains were repeatedly obtained at different distances from the Sun, including the aphelion region of its orbit that is favorable for studies of the nucleus with very little coma contamination. These observations indicate that Encke is a gas-rich comet, i.e. its continuum is extremely weak with respect to the molecular emissions with dust-to-gas mass ratio <0.1 in the visible (Newburn and Spinrad, 1985). Gehrz et al. (1989) found that in the thermal infrared Encke showed very little continuum superheat and a weak silicate emission feature. On the other hand, from the ISO-CAM observations in the thermal infrared domain, Reach et al. (2000) concluded that the dust-to-gas mass ratio of comet Encke was 10–30, much higher than found from the visible light observations. They concluded that the coma of Encke contained particles with radii as large as 5–10 μm. Using the mid- and far-infrared data, Lisse et al. (2004) confirmed that comet Encke emits a much higher fraction of large grains, with the sizes 20 μm and higher. A dust trail extending along the orbit was detected in Encke from infrared observations (Sykes and Walker, 1992). The existence of the dust trail indicates that the comet also releases a large number of millimeter and centimeter sized particles (Reach et al., 2000). Thus, comet Encke looks as the dust-poor comet in the visible because the fraction of particles with size comparable to the wavelength (submicron and micron-sized) is depleted in comparison to other comets (A'Hearn et al., 1995).

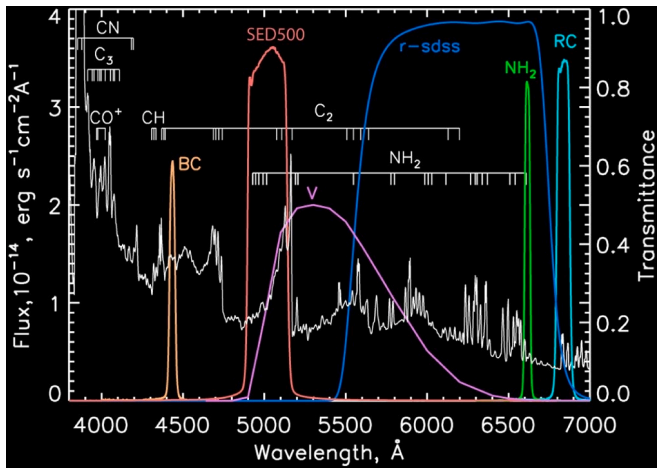

**Fig. 1.** Normalized transmission curves of the used filters BC, RC, NH$_2$, SED500, r-sdss, and V superimposed on the observed spectrum of comet 2P/Encke on January 23, 2017.

The dust coma of comet Encke is concentrated in a small near-nucleus region whereas molecular emissions are extended to large distances from the nucleus (Jockers et al., 2005). The abundance of CN, C$_2$, and C$_3$ molecules is typical while NH is depleted by a factor of 2–3 (Osip et al., 1992). Color of the comet is "reddish" in the domain from λ4000 to 6900 Å (Newburn and Spinrad, 1985).

Comet Encke shows a fan directed toward the Sun which was observed near the perihelion in almost all its appearances. This type of coma structure assumes that gas and dust outflow from the nucleus surface only on the illuminated side (Festou and Barale, 2000). Sometimes there were one or two jets in the comet indicating the presence of small dust particles.

Fernández et al. (2000, 2014) obtained the following physical, optical, and rotational characteristics of the nucleus of Encke: an effective radius is of 2.4 ± 0.3 km with one axial ratio at least 2.6, a geometric albedo in the R band is of 0.047 ± 0.023, a linear slope of photometric phase function is 0.06 mag/deg. Colors were found to be V − R = 0.39 ± 0.06 and B − V = 0.73 ± 0.06 (Lowry and Weissman, 2007) that is consistent with the spectral slope $S$ found by Jewitt (2002) to be 8.9 ± 1.6%/1000 Å. The rotation period of the comet nucleus was measured by different methods and was found to be on average of 11.1 h (Belton et al., 2005; Fernández et al., 2005; Harmon and Nolan, 2005; Lowry and Weissman, 2007; Jockers et al., 2011). According to Sekanina (1979) and Reach et al. (2000), the spin axis of the nucleus of Encke lies nearly in its orbital plane and an active source close to the pole produces the Encke's sunward fan that results in strong seasonal variations of the particle emission.

Levison et al. (2006) suggests that Encke had been dormant for most of its dynamical life; however its activity resumed as its perihelion distance decreased. Approaching the perihelion, the comet shows an activity that is manifested by the appearance of a fan-shaped coma. Sarugaku et al. (2015) have proposed that Encke's nucleus surface is thermally evolved and formation of the dust mantle has progressed. According to the hypothesis of Newburn and Spinrad (1985), the amount of gas diffusing through the mantle increases as the comet approaches the Sun, while the ratio of the dust and gas, which flow from a major active area, is roughly constant, resulting in the rise of the gas-to-dust production rate. Despite a low brightness, Encke is rather active comet: at perihelion, it releases >$10^3$ kg s$^{-1}$ of dust and is considered as one of the main sources of zodiacal particles and the Taurid meteoroid complex (Epifani et al., 2001).

In this paper, we present results of our observations of comet Encke in the 2013 and 2017 apparitions. In Section 2, we describe our photometric and spectral observations of the comet and data reduction. The results of spectral observations are analyzed in Section 3. A description of morphological features of the coma is given in Sections 4 and 5. The gas and dust production rates obtained from the photometric and spectral observations along with the comparison with earlier works are given in Section 6. The surface brightness profiles across the dust coma are analyzed in Section 7, and contribution of the nucleus to the total brightness is determined in Section 8. Color map and normalized reflectivity are derived in Section 9. In Section 10, the spatial distribution of the dust/gas ratio over the coma is analyzed. A dynamical simulation of morphological features of the coma is described in Section 11. We discuss our results in Section 12 and give our conclusions in Section 13.

## Observations and data reduction

*Instrument and general features*

Observations of comet Encke were carried out at the 6-m BTA telescope of the Special Astrophysical Observatory (Russia) on November 4, 2013 (17 days before the perihelion passage of the comet on November 21, 2013) and January 23, 2017 (46 days before the perihelion of the comet on March 10, 2017), when the comet was located at heliocentric distance of 0.56 and 1.05 au, geocentric distance of 0.65 and 1.34 au, and phase angle was about 109.2° and 46.8°, respectively. A focal reducer SCORPIO-2 (Spectral Camera with Optical Reducer for Photometrical and Interferometrical Observations) installed at the primary focus of the telescope was used in the photometric and spectroscopic modes (Afanasiev and Moiseev, 2011; Afanasiev and Amirkhanyan, 2012). The back-illuminated CCD detector EEV 42–90 consisting of 2048 × 2048 pixels with a pixel size of 13.5 × 13.5 μm was used. The full field of view was 6.1′ × 6.1′ with a pixel scale of 0.18 arcsec/px.

The telescope tracked the motion of the comet to compensate its proper velocity during the exposures. As a result, we received direct images (hereafter Ima) and long-slit spectra (Sp). Binning 2 × 2 was applied to the photometric images and 2 × 1 was applied to the spectroscopic frames to improve the signal/noise (S/N) ratio.

Unfortunately, the poor weather conditions (cloudy) limited us to only one night in every set of observations, preventing the implementation of the planned observing program of comet Encke during both periods. Only a few observations were used for the analysis of comet morphology in November of 2013 due to non-photometric conditions. In January 23, 2017, the observations were obtained during excellent sky conditions. It was a single truly clear night for our observing period at the telescope. The seeing (*FWHM*) was 2.5″ (1186 km in November 2013) and 1.1″–1.2″ (1114 km in January 2017). Primary reduction of the data was performed using the IDL codes developed at the SAO RAS, detailed description of the image processing, reduction, errors estimation are similar to those described by Afanasiev and Moiseev (2011), Afanasiev and Amirkhanyan (2012), Ivanova et al. (2015, 2017a, 2017b) where also more details on the instrument can be found.

**Table 1**
Log of the observations of comet 2P/Encke.

| Date, UT | r (au) | Δ (au) | α (deg) | D (km/px) | Filter/grism | $T_{exp}$ (sec) | N | Mode |
|---|---|---|---|---|---|---|---|---|
| 2013 November 4 | | | | | | | | |
| 02:50:40–02:56:07 | 0.561 | 0.654 | 109.2 | 85.4 | V | 70 | 7 | Ima |
| 2017 January 23 | | | | | | | | |
| 15:08:50–15:14:57 | 1.051 | 1.336 | 46.8 | 174.4 | r-sdss | 40 | 2 | Ima |
| 15:17:25–15:36:42 | 1.051 | 1.336 | 46.8 | 174.4 | SED500 | 270 | 5 | Ima |
| 15:22:26–15:34:06 | 1.051 | 1.336 | 46.8 | 174.4 | RC | 180 | 3 | Ima |
| 15:23:52–15:35:23 | 1.051 | 1.336 | 46.8 | 174.4 | $NH_2$ | 180 | 3 | Ima |
| 15:26:25–15:38:00 | 1.051 | 1.336 | 46.8 | 174.4 | BC | 180 | 3 | Ima |
| 16:31:24–16:35:33 | 1.050 | 1.335 | 46.8 | 174.4 | V | 30 | 3 | Ima |
| 16:40:54–17:00:47 | 1.049 | 1.335 | 46.8 | – | VPHG1200@540 | 940 | 6 | Sp |

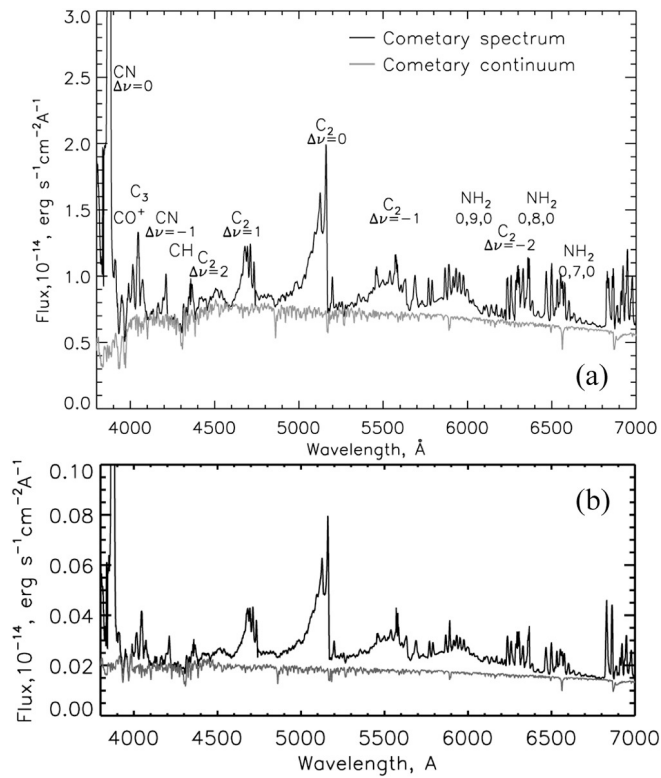

**Fig. 2.** Spectrum of comet 2P/Encke with the scaled solar spectrum (Neckel and Labs, 1984) obtained on January 23.82, 2017: (a) – the spectrograph slit extracted from the coma the projected area of 355,000 km × 969 km; (b) – the spectrum of coma area between 5000 km and 10,000 km from the optocenter, excluding the near-nucleus region.

*Photometric observations*

On November 4, 2013, direct images of comet Encke were only obtained with the broad-band Johnson-Cousins V filter (we represent the central wavelength $\lambda_0$ and *FWHM* as λ5580/880 Å). The narrow-band cometary filters BC (λ4429/36 Å), RC (λ6835/83 Å), and $NH_2$ (λ6615/34 Å) from the ESA filter set as well as medium-band filter SED500 (λ5019/246 Å) and broad-band V (λ5580/880 Å) and r-sdss (λ6200/1200 Å) filters were used for observations obtained on January 23, 2017. The transmission curves of the used filters BC, RC, SED500, r-sdss, and V superimposed on the observed spectrum of comet Encke are shown in Fig. 1.

The night was photometric allowing absolute calibration of the photometric data. For the photometric calibration of the images,

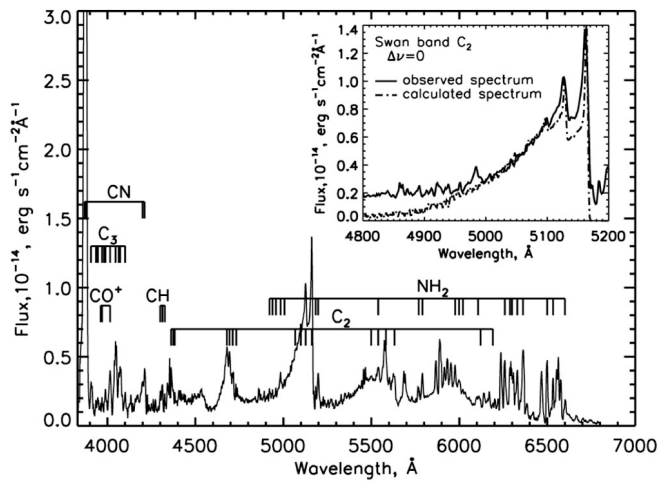

**Fig. 3.** Emission spectrum of comet 2P/Encke with the main identified species. The insert shows a zoom-in of a comparison of the observed spectrum (solid line) and the modeled spectrum (dash line) of the $C_2(\Delta \nu = 0)$ Swan band.

standard star HD 52266 (Farnham et al., 2000) was observed. The spectral behavior of the atmospheric transparency was taken from Kartasheva and Chunakova (1978).

The reduction process of the raw data included bias subtraction, flat-fielding, and removing traces of cosmic rays. The bias was removed by subtracting an averaged frame with zero exposure time. The morning sky was exposed to provide flat-field corrections for the non-uniform sensitivity of the CCD matrix. Removal of the traces of cosmic rays was done at the final stage of the reduction via a robust parameter estimates for reducing a bias caused by outliers (Fujisawa, 2013). We used the sky background level in the regions of the image, which were not covered by the cometary coma and free of faint stars, based on estimations from histogram of counts in the image. To increase the S/N ratio, we added together all images of the comet taken during night and summed using a robust averaging method (Rousseeuw and Bassett, 1990). To avoid artifacts, the images of the comet were centered with a precision of 0.1 px using the central contour of relative intensity (isophotes) which was closest to the maximum brightness of the comet.

In order to segregate low-contrast structures in the images, we used various digital processing techniques: a rotational gradient method (Larson and Sekanina, 1984), Gauss blurring, division by azimuthal average and median filtering (Samarasinha and Larson, 2014). To exclude spurious features when interpreting the obtained images, each of the digital filters was applied with the same processing parameters to all individual exposures of the comet as well as to the composite frames, as this was done by Manzini et al. (2007). This technique was already used to pick out structures in several comets with good results (Ivanova et al., 2009, 2017b; Rosenbush et al., 2017).

**Table 2**
Emissions in the spectrum of comet 2P/Encke.

| $\lambda_{obs}$ (Å) | $F_{obs} \times 10^{-15}$ (erg s$^{-1}$ cm$^{-2}$ Å$^{-1}$) | $\lambda_{ref}$ (Å) | Molecule | Electron transition | Vibrational transition |
|---|---|---|---|---|---|
| 3781.88 | 14.67 | 3785.89 | CO$^+$ | A$^2\Pi_i$–X$^2\Sigma^+$ | 4–0 |
| 3798.08 | 16.19 | 3802.41 | C$_3$ | A$^1\Pi_u$–X$^1\Sigma^1_g$ | |
|  |  | 3803.25 | CO$^+$ | A$^2\Pi_i$–X$^2\Sigma^+$ | 4–0 |
| 3825.62 | 9.26 | 3826.08 | C$_3$ | A$^1\Pi_u$–X$^1\Sigma^1_g$ | |
| 3869.36 | 57.22 | 3869.00 | CN | B$^2\Sigma^+$–X$^2\Sigma^+$ | 0–0 |
| 3879.08 | 72.25 | 3878.90 | CN | B$^2\Sigma^+$–X$^2\Sigma^+$ | 0–0 |
| 3906.62 | 20.09 | 3901.84 | C$_3$ | A$^1\Pi_u$–X$^1\Sigma^1_g$ | |
|  |  | 3914.47 |  |  |  |
| 3934.16 | 16.55 | 3925.52 | C$_3$ | A$^1\Pi_u$–X$^1\Sigma^1_g$ | |
|  |  | 3936.56 |  |  |  |
| 3945.50 | 15.88 | 3942.88 | C$_3$ | A$^1\Pi_u$–X$^1\Sigma^1_g$ | |
|  |  | 3949.19 |  |  |  |
| 3969.80 | 14.30 | 3964.97 | C$_3$ | A$^1\Pi_u$–X$^1\Sigma^1_g$ | |
|  |  | 3971.29 |  |  |  |
| 3981.14 | 13.27 | 3983.12 | C$_3$ | A$^1\Pi_u$–X$^1\Sigma^1_g$ | |
| 3989.24 | 13.84 | 3990.23 | C$_3$ | A$^1\Pi_u$–X$^1\Sigma^1_g$ | |
| 4015.16 |  | 4005.28 | CO$^+$ | A$^2\Pi_i$–X$^2\Sigma^+$ | 4–0 |
|  | 14.00 | 4025.79 |  |  |  |
|  |  | 4012.32 |  |  |  |
|  |  | 4018.64 | C$_3$ | A$^1\Pi_u$–X$^1\Sigma^1_g$ | |
| 4045.94 | 19.93 | 4039.16 | C$_3$ | A$^1\Pi_u$–X$^1\Sigma^1_g$ | |
|  |  | 4050.21 |  |  |  |
| 4063.76 | 11.85 | 4062.83 | C$_3$ | A$^1\Pi_u$–X$^1\Sigma^1_g$ | |
| 4071.86 | 12.05 | 4072.30 | C$_3$ | A$^1\Pi_u$–X$^1\Sigma^1_g$ | |
| 4101.02 | 9.29 | 4099.13 | C$_3$ | A$^1\Pi_u$–X$^1\Sigma^1_g$ | |
| 4199.84 | 8.09 | 4202.90 | CN | B$^2\Sigma^+$–X$^2\Sigma^+$ | 0–1 |
| 4211.18 | 8.94 | 4212.80 | CN | B$^2\Sigma^+$–X$^2\Sigma^+$ | 0–1 |
| 4300.28 | 6.95 | 4303.58 | CH | A$^2\Delta$–X$^2\Pi$ | 0–0 |
| 4313.24 | 7.35 | 4316.21 | CH | A$^2\Delta$–X$^2\Pi$ | 0–0 |
| 4326.20 | 6.71 | 4333.57 | CH | A$^2\Delta$–X$^2\Pi$ | 0–0 |
| 4361.84 | 12.22 | 4363.79 | C$_2$ | A$^3\Pi_g$–X$^3\Pi_u$ | $\Delta v=+2$ |
| 4374.80 | 6.98 | 4370.13 | C$_2$ | A$^3\Pi_g$–X$^3\Pi_u$ | $\Delta v=+2$ |
| 4382.90 | 6.80 | 4381.23 | C$_2$ | A$^3\Pi_g$–X$^3\Pi_u$ | $\Delta v=+2$ |
| 4679.36 | 8.65 | 4677.69 | C$_2$ | A$^3\Pi_g$–X$^3\Pi_u$ | $\Delta v=+1$ |
|  |  | 4684.04 |  |  |  |
| 4693.94 | 8.13 | 4696.72 | C$_2$ | A$^3\Pi_g$–X$^3\Pi_u$ | $\Delta v=+1$ |
| 4711.76 | 8.10 | 4714.16 | C$_2$ | A$^3\Pi_g$–X$^3\Pi_u$ | $\Delta v=+1$ |
| 4732.82 | 6.38 | 4736.35 | C$_2$ | A$^3\Pi_g$–X$^3\Pi_u$ | $\Delta v=+1$ |
| 4920.74 | 2.71 | 4925.01 | NH$_2$ | A$^2A^1$–X$^2B^1$ | (0,13,0) |
| 4938.56 | 2.64 | 4937.70 | NH$_2$ | A$^2A^1$–X$^2B^1$ | (0,13,0) |
| 4958.00 | 2.76 | 4955.14 | NH$_2$ | A$^2A^1$–X$^2B^1$ | (0,13,0) |
| 4983.92 | 3.90 | 4980.50 | NH$_2$ | A$^2A^1$–X$^2B^1$ | (0,13,0) |
|  |  | 4993.19 |  |  |  |
| 5006.60 | 3.38 | 5007.45 | NH$_2$ | A$^2A^1$–X$^2B^1$ | (0,13,0) |
| 5067.56 | 5.48 | 5069.28 | C$_2$ | A$^3\Pi_g$–X$^3\Pi_u$ | $\Delta v=0$ |
| 5098.94 | 7.39 | 5096.24 | C$_2$ | A$^3\Pi_g$–X$^3\Pi_u$ | $\Delta v=0$ |
| 5126.48 | 10.32 | 5127.94 | C$_2$ | A$^3\Pi_g$–X$^3\Pi_u$ | $\Delta v=0$ |
| 5162.12 | 13.63 | 5164.41 | C$_2$ | A$^3\Pi_g$–X$^3\Pi_u$ | $\Delta v=0$ |
| 5183.18 | 2.88 | 5186.60 | NH$_2$ | A$^2A^1$–X$^2B^1$ | (0,12,0) |
| 5197 | 3.82 | 5194.53 | NH$_2$ | A$^2A^1$–X$^2B^1$ | (0,12,0) |
|  |  | 5207.21 |  |  |  |
| 5498.85 | 3.57 | 5500.51 | C$_2$ | A$^3\Pi_g$–X$^3\Pi_u$ | $\Delta v=-1$ |
| 5539.58 | 4.25 | 5540.15 | C$_2$ | A$^3\Pi_g$–X$^3\Pi_u$ | $\Delta v=-1$ |
|  |  | 5540.15 | NH$_2$ | A$^2A^1$–X$^2B^1$ | (0,11,0) |
| 5583.32 | 7.47 | 5584.54 | C$_2$ | A$^3\Pi_g$–X$^3\Pi_u$ | $\Delta v=-1$ |
| 5631.92 | 3.62 | 5633.68 | C$_2$ | A$^3\Pi_g$–X$^3\Pi_u$ | $\Delta v=-1$ |
| 5769.62 | 3.77 | 5767.45 | NH$_2$ | A$^2A^1$–X$^2B^1$ | (0,10,0) |
| 5790.68 | 383 | 5789.73 | NH$_2$ | A$^2A^1$–X$^2B^1$ | (0,10,0) |
| 5976.98 | 4.37 | 5976.06 | NH$_2$ | A$^2A^1$–X$^2B^1$ | (0,9,0) |
| 5998.04 | 3.31 | 5995.18 | NH$_2$ | A$^2A^1$–X$^2B^1$ | (0,9,0) |
|  |  | 6006.33 |  |  |  |
| 6020.72 | 2.65 | 6019.09 | NH$_2$ | A$^2A^1$–X$^2B^1$ | (0,9,0) |
| 6107.09 | 2.06 | 6097.21 | NH$_2$ | A$^2A^1$–X$^2B^1$ | (0,9,0) |
|  |  | 6109.97 |  |  |  |
|  |  | 6121.13 |  |  |  |
| 6121.16 | 1.87 | 6121.13 | C$_2$ | A$^3\Pi_g$–X$^3\Pi_u$ | $\Delta v=-2$ |
| 6190.82 | 1.67 | 6189.74 | C$_2$ | A$^3\Pi_g$–X$^3\Pi_u$ | $\Delta v=-2$ |
| 6258.86 | 4.75 | 6259.97 | NH$_2$ | A$^2A^1$–X$^2B^1$ | (0,9,0) |
|  |  | 6267.95 |  |  |  |
| 6288.02 | 4.25 | 6285.52 | NH$_2$ | A$^2A^1$–X$^2B^1$ | (0,8,0) |
| 6299.36 | 14.23 | 6298.29 | NH$_2$ | A$^2A^1$–X$^2B^1$ | (0,8,0) |
| 6330.14 | 4.04 | 6331.83 | NH$_2$ | A$^2A^1$–X$^2B^1$ | (0,8,0) |
| 6362.54 | 14.74 | 6360.58 | NH$_2$ | A$^2A^1$–X$^2B^1$ | (0,8,0) |
| 6500.24 | 4.74 | 6501.24 | NH$_2$ | A$^2A^1$–X$^2B^1$ | (0,8,0) |
| 6532.64 | 3.47 | 6533.22 | NH$_2$ | A$^2A^1$–X$^2B^1$ | (0,8,0) |
|  |  | 6536.42 |  |  |  |
| 6600.68 | 1.92 | 6600.42 | NH$_2$ | A$^2A^1$–X$^2B^1$ | (0,7,0) |

## 2.3. Spectral observations

The long-slit spectroscopy of comet Encke was performed on January 23, 2017 with the transparent grism VPHG1200@540 which operated in the wavelength range of 3600–7000 Å. The spectrograph slit with 6.1′ × 1.0″ dimensions (354,654 km × 969 km at the comet) was centered on the photometric nucleus (i.e., the brightest point of the coma) and oriented across the cometary coma in the direction of the velocity vector of the comet at position angle of 74.6°. The obtained spectrum has the spectral resolution of about 5 Å across the whole wavelength range. The raw spectrum of comet Encke is shown in Fig. 1, in which the strongest emissions are indicated. The data have been treated using a standard reduction procedure, as bias subtraction, flat field correction, removal of cosmic ray traces, spectral line curvature compensation, wavelength calibration, and flux calibration, followed by removal of the sky spectrum and the solar continuum. For determination of the flat-field correction, we used the spectrum of a built-in lamp with a continuous spectrum. The wavelength calibration of the spectrum was performed using the line spectrum emitted by a He-Ne-Ar lamp. All spectra were stacked and combined with the robust averaging algorithm, in order to remove cosmic ray traces and improve the S/N ratio. Observations of standard star BD33°2642 from Oke (1990) were recorded at the same night for absolute flux calibration of the cometary spectrum. Since the comet moved during the observations with a speed of 27.98 km/s, the Doppler shift was taken into account: the appropriate coefficient was 1.00009.

The viewing geometry and log of observations of comet Encke for the two periods are presented in Table 1. There we list the date of observation and the range of UT for each type of observation, the heliocentric ($r$) and geocentric ($\Delta$) distances, the phase angle (Sun-Comet-Earth angle) ($\alpha$), the pixel size at the distance of the comet ($D$), the filter or grism, the total exposure time during the night ($T_{exp}$), number of cycles of exposures obtained in one night ($N$), and the mode of the observation. As we wrote above, our observations of comet Encke in November of 2013 were extremely limited. Weather conditions were very poor over the night of our observations, resulting in images useful only for studying coma morphology in the V filter.

## Analysis of observed spectra

In Fig. 2a, we show a composite spectrum of coma area of 355,000 km × 969 km in comet Encke, demonstrating its strong emission activity. For comparison of the relative brightness of the emission bands and underlying continuum, the spectrum of the area between 5000 and 10,000 km from the optocenter is represented in Fig. 2b. To isolate the emission spectrum, we subtracted fitted continuum from the observed spectrum. For this, the spectrum of the Sun (Neckel and Labs, 1984) was reduced to the cometary spectrum resolution by the Gaussian convolution having the corresponding halfwidth of the profile. The solar spectrum was shifted in such a way that its level was as close as possible to the lower boundary of the cometary spectrum in the spectral windows where the continuum dominated. The emission spectrum and continuum are presented in Fig. 2a, in which the main identified emissions are labeled. To obtain the pure emission spectrum of the comet (Fig. 3), the continuum was removed from the total cometary spectrum. Most known emissions in optical domain were detected in the spectrum. We detected 60 features belonging to CN, C$_3$, C$_2$, CH, NH$_2$, and CO$^+$ molecules. Results of identification are listed in Table 2, where the following columns are presented: the wavelengths $\lambda_{obs}$ and $\lambda_{ref}$, emission intensity, molecule, electron and vibrational transitions. $\lambda_{ref}$ is a wavelength from the reference molecular spectrum that forms an emission feature in the observed cometary spectrum. In the case of several lines of the same molecule, we used the symbol "{".

For identification of the emissions in the cometary spectrum, theoretical and laboratory spectra of the molecules, observed in comets, were used. To compare the observed and calculated spectra, the latter were reduced to the spectral resolution of the observations by the Gaussian convolution having the corresponding halfwidth. Phillips and Davis (1968), Luque and Crosley (1999), Gausset et al. (1965), Dressler and Ramsay (1959), Dobrovolsky (1966), and Kim (1994) were used as reference sources for CN, $C_2$, $C_3$, $NH_2$, CH, and $CO^+$ molecules, respectively.

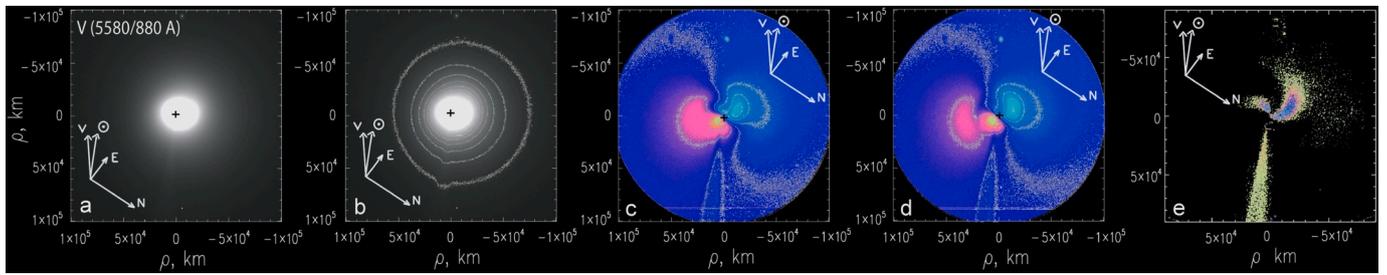

**Fig. 4.** Image of comet 2P/Encke taken through the V filter on November 4.08, 2013: (a) shows the composite image of the comet; (b) shows the relative isophotes; (c), (d), and (e) are intensity images processed by the digital filters, respectively, division by azimuthal average, azimuthal renormalization, and rotational gradient method (Larson and Sekanina, 1984; Samarasinha and Larson, 2014). The isophotes are superimposed on the processed images c and d. Arrows point the direction to the Sun (⊙), North (N), East (E), and the heliocentric velocity vector of the comet as projected onto the plane of the sky (V). Negative distance means the direction to the Sun, and positive distance is in the antisolar direction.

Initially, asymmetry in the spatial profile of molecular emissions in the head of the comet Encke was observed by Swings and Haser (1956) in the 1937 and 1947 apparitions. Our observations of comet Encke in 2017 and 2003 (Shubina et al., 2018) showed similar asymmetry in emissions. Intensity of $C_3$ emission detected in the spectrum of the comet (see Fig. 3) is higher than that of CN (0–1) emission. The ratio of $C_3$ ($\lambda 4313$ Å) to CN (0–1) is larger than 1 that is typical for the comets observed at heliocentric distances larger than 1 au (Dobrovolsky, 1966).

**Photometry: observations on November 4, 2013**

Fig. 4 shows a V-band image of comet Encke obtained on November 4, 2013, i.e. 17 days before the perihelion passage of the comet on November 21, 2013. Here, we depict the averaged composite image of the comet (a) and relative isophotes differing by a factor $\sqrt{2}$ (b). The image is the robust average of seven original exposures of 10 s. As it is seen in the figure, the comet is fairly well condensed; the bright and nearly symmetric coma is ~20,000 km across, although a weak elongation of the coma is visible in the sunward direction. Note that the optocenter of the comet (marked by a white cross), defined by the central isophote, does not coincide with the center of symmetry of the bright coma and is shifted in the antisolar direction. The comet exhibits a faint straight-line structure, which can be seen across the image in the direction away from the Sun, $PA = 312.7°$ (the direction to the Sun is 128.9°). As it is visible from the isophotes (panel b), this feature expands with the distance from the nucleus.

To reveal the low-contrast structures in the cometary coma, we applied some available image enhancement techniques: division by azimuthal average (panel (c)); azimuthal renormalization (d) (Samarasinha and Larson, 2014); and rotational gradient method (e) (Larson and Sekanina, 1984). Two first techniques allow removing the bright background from the cometary coma and highlighting the low-contrast features. The division by azimuthal average is very good to separate brighter broad jets. There is a number of evidences that coma morphology does not contain artifacts and it is not modulated by rotation. Images were obtained at very short time intervals (see Table 1) as compared with the rotation period of the nucleus that is about 11.1 h. Processing images by digital filters was performed by two different software: one from the Samarasinha's website (http://www.psi.edu/research/cometimen) and our own MATLAB codes for all these filters. As we indicated above, intensity maps were constructed by stacking together all the photometric images taken with the same filter. We have applied a combination of numerical techniques and visual inspection. Since the different enhancement techniques affect the image in different ways and to exclude spurious features after the images have been enhanced, we applied each of the digital filters to each individual frame as well as to the same composite image (see e.g., Manzini et al., 2007) to evaluate whether revealed features were real or not. This technique allowed us to avoid appearance of false features when processing the images by digital filters. We also studied the change in the jet structure due to the shift of comet's optocenter. Even in the case of significant displacement of the optocenter by 6 px, the jet structure was preserved.

The processing images revealed complex coma morphology. As one can see in Fig. 4, images (c) and (d), the most prominent large-scale features are large twisted structures. The bright complex structures observed in the coma could be considered as a result of outflow of matter

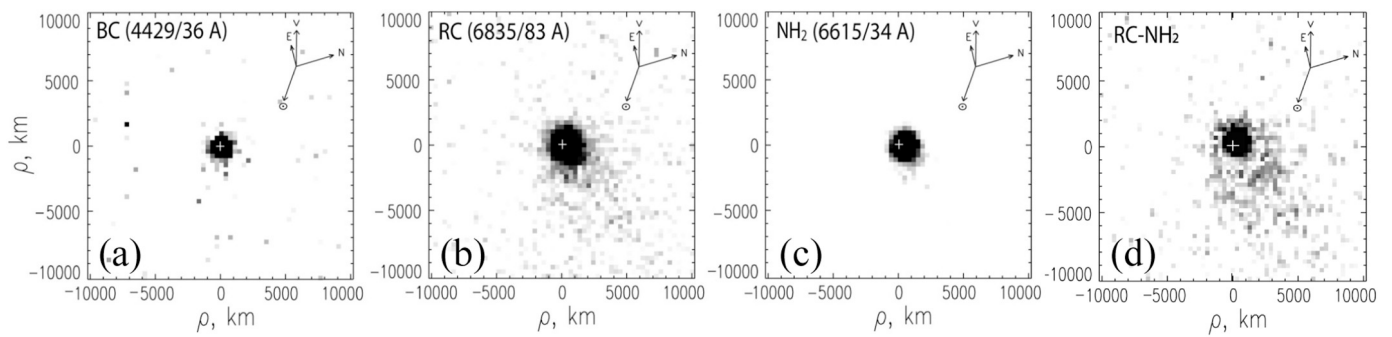

**Fig. 5.** Images of comet 2P/Encke obtained through the narrow-band blue continuum BC (a) and red continuum RC (b) filters and the filter centered on the band system of $NH_2$ (c) on January 23.75, 2017. Dust intensity map obtained after subtraction of the $NH_2$ intensity image from the RC image is denoted as RC′ and shown on plot (d). The optocenter is marked with a white cross. Positive distance is in the antisolar direction, and negative distance is in the solar direction. North, East, sunward, and velocity vector directions are indicated. (For interpretation of the references to color in this figure legend, the reader is referred to the web version of this article.)

# Filter SED500

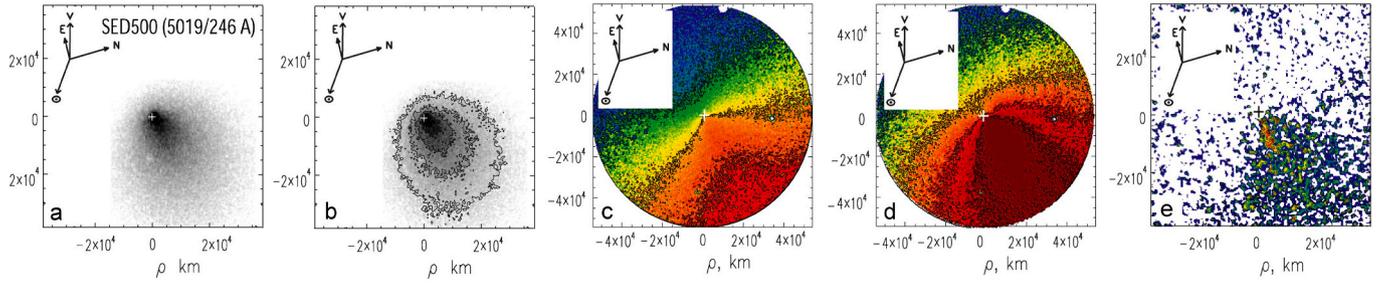

# Filter V

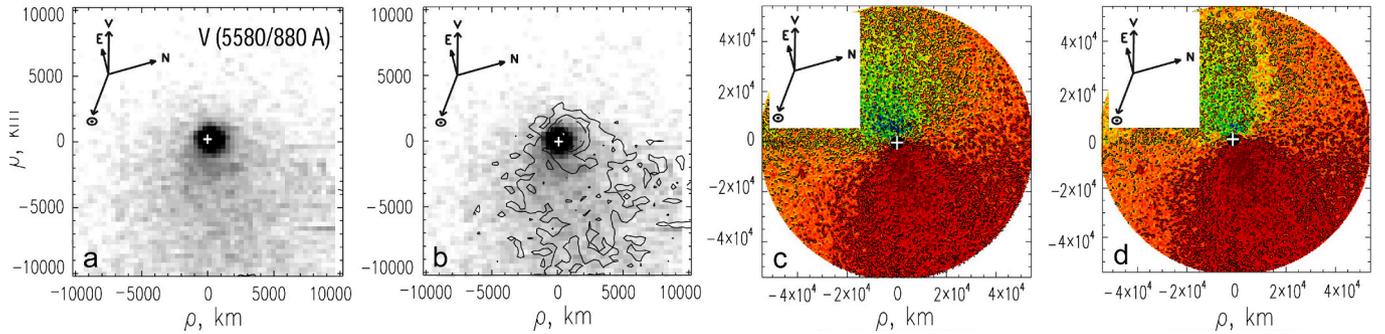

**Fig. 6.** Images of comet 2P/Encke in the SED500 (upper row), V (middle row), and r-sdss (bottom row) filters obtained on January 23, 2017. (a) shows the direct images of the comet; (b) is the relative isophotes differing by a factor $\sqrt{2}$; (c) and (d) represent intensity images to which we applied division by azimuthal average and division by azimuthal median, respectively (Samarasinha and Larson, 2014); (e) is intensity image processed by a rotational gradient method (Larson and Sekanina, 1984; Samarasinha and Larson, 2014). The red rectangle shows the location of the spectrograph slit which extracted from the coma the projected area of 355,000 km × 969 km. North, East, sunward, and velocity vector are indicated in each frame. (For interpretation of the references to color in this figure legend, the reader is referred to the web version of this article.)

from some active region located near the subsolar point of the rotating nucleus. To enhance possible anisotropies in the coma structures, the image was processed by rotational gradient method. As one can see in panel (e), the applied enhancement procedures resulted in detection of a highly structured coma with a plasma tail and two "wing-like" structures (jets?) which are strongly curved due to a rotation of the comet nucleus and separation of particles by mass. Our conclusion that the linear feature is the plasma tail primarily based on visual inspection. The observed tail is straight and located nearly in the antisunward direction. A slight deviation from this direction is caused by the geometry of observations and nucleus spin axis orientation. The position angles of the jet-like features are difficult to determine precisely, given their large apparent curvature. Therefore, we used the directions the jets have near the nucleus: $PA = 42.8°$ for the strong jet extending to the right (jet 1) and $PA = 173.5°$ for the small jet (jet 2) pointing to the left. However, as it will be shown in Section 11, in fact there is only one jet.

### Photometry: observations on January 23, 2017

*Observations with narrow-band filters*

In January of 2017, i.e. 46 days before the perihelion passage of the comet on March 10, 2017, comet Encke exhibited a very different appearance compared with the observations of 2013. Fig. 5 shows three narrow-band images of comet Encke taken with the cometary contin-uum BC ($\lambda 4429/36$ Å) and RC ($\lambda 6835/83$ Å) filters designed to avoid a great contribution of gas emissions and in the filter ($\lambda 6615/34$ Å) centered on the (0,7,0) transition of the $\tilde{A}_2 A_1 - \tilde{X}_2 B_1$ electronic band system of the $NH_2$ molecule. The comet in the BC filter displays a sharp, round central condensation, perhaps surrounded by a very weak small coma, i.e. almost the entire observed coma is within the seeing disc. This indicates that the dust in comet Encke is mainly concentrated near the nucleus. The RC image also exhibits a central condensation, however much larger in diameter than that in the BC filter (>2000 km) and slightly elongated at the position angle 298°. There is also a dust coma elongated to the distances 7000–10,000 km. This elongation does not exactly coincide with the direction to the Sun. It seems that this feature is a fan usually observed in comet Encke in the solar hemisphere.

The $NH_2$ image of comet Encke is slightly asymmetric with respect to the optocenter and slightly elongated into the fan. Since the $NH_2$ emission can penetrate the RC band, we subtracted the coadded $NH_2$ images of the comet from the coadded RC image. For correct subtracting, we took into account the different light transmission of the filters and the different continuum levels from our spectral data. We also took into account the reddening gradient $k = 1.308$ determined from the spectrum. The images of the comet in two filters were centered using the central isophote, which was closest to the maximum brightness of the comet. Fig. 5(d) shows the intensity map of pure continuum after we subtracted the $NH_2$ intensity from the observed intensity in the RC band. In this image, denoted as RC′, we clearly detect sunlight scattered from dust particles distributed in the sunward direction to at least 10,000 km. Comparing plots (c) and (d), we can conclude that the dust contributes significantly to the surface brightness of the fan.

### 5.2. Coma morphology with broadband filters

Fig. 6 illustrates the direct images of comet Encke acquired in the broadband filters, relative isophotes, and the images processed by different digital techniques. The upper row shows the image of the comet in the middle-band filter SED500 taken on January 23, 2017. The middle and bottom rows display the comet in the V and r-sdss filters, respectively, obtained on the same night. Panels (c) and (d) represent the processed intensity images to which we applied division by azimuthal average and division by azimuthal median, respectively (Samarasinha and Larson, 2014). The panel (e) is the intensity image processed by a rotational gradient method (Larson and Sekanina, 1984; Samarasinha and Larson, 2014). The comet in the V and r-sdss filters is rather faint and therefore this method gives poor results. Comparison of Figs. 5 and 6 shows that the morphology of the comet is similar in the broadband and dust continuum images, but in the latter, the coma produced by the dust, looks very faint. Since our other results show concentration of the dust in the near-nucleus coma, we presume that the central isophotes are associated with the sunlight reflected by solid particles.

As it can be seen in Fig. 6, all images show the same morphology of the coma: a prominent fan is observed in the sunward sector, but not directly centered on the sunward vector, in all photometrical bands. This type of coma structure assumes that gas and dust were released from the sunward side of the nucleus. According to Sekanina (1979), cometary sunward fans are cones of emission produced when a high-latitude source on the Sun-facing hemisphere is carried around the rotational pole.

A rotational gradient enhancement technique applied to the SED500 images was required to reveal the presence of a fan-like structure and the "short jet" coming from the near-nucleus region (panel e). An extension of this feature in the coma is about 7000 km of projected nucleus distance, the position angle is 289.6°, and it coincides with an axis of symmetry of the projected fan. Such structure is also seen in the images obtained through the V and r-sdss filters (see c and d panels in the middle and bottom rows). The "short jet" is the brightest part of the near-sunward fan, located near its axis of symmetry. It is likely that this narrow "short jet" indicates a localized coma feature of enhanced dust reflectivity produced by some initial ejection of the matter from the nucleus.

Prominent sunward-facing fan of comet Encke was visible on nearly every recorded apparition since 1896 (Sekanina, 1988a, 1988b). According to this author, the fan-shaped coma of comet Encke, observed along the inbound part of the comet's orbit, is associated with outgassing from an active region located in the north hemisphere at the latitude 55°. At the same time, the fan-like feature observed after perihelion, i.e. along outbound part of the orbit, is a product of emission from a second active region located in the south hemisphere at the latitude −75°. If the activity of the nucleus was restricted to one small active region, one would expect not only a change in the absolute level of the nucleus output due to changing insolation conditions coupled to nucleus rotation, but also a change in the orientation of the fan (Festou and Barale, 2000). This may indicate the presence of several small active regions on a rotating nucleus.

The axis of rotation in our case lies almost in the picture plane. Therefore, we see the substance emission almost from the side. The rotation of the nucleus leads to the formation of a cone of ejected substance. As a result, we see the fan on the line of sight. Festou and Barale found that for all viewing geometries, the mean orientation of the projected fan axis is 15° from the sunward direction and for most of the time the fan width is close to 80°. However, according to our measurements, the angle between the sunward direction ($PA = 239.1°$) and the projected fan axis ($PA = 289.6°$) is about 50°, and the fan width is within the range from ∼60° to ∼80° depending on the filter.

In their article, Festou and Barale (2000) noted that H. Spinrad and S. Djorgovski observed a wide fan in comet Encke on October 10, 1980, whose axis deviated from the direction to the Sun by 10°–20°. Using our model (see Section 11 in detail), we calculated possible configurations of jets for observations of 1980 and 2017. For this, two sets of parameters were considered and analyzed: i) the coordinates of the north pole of the comet spin axis $RA = 200°$ and $Dec = +5°$ and the cometocentric latitude of the active region $\varphi = +55°$ for our observations; ii) the parameters which were used by Festou and Barale (2000) for observations of 1980, $RA = 200°$ and $Dec = +1.7°$, $\varphi = +50°$. We found that the deviation of the fan axis by ∼15° from the direction to the Sun on October 10, 1980 and ∼50° on January 23, 2017 may be explained by the difference in the positions of the Earth and the comet with respect to the Sun.

## 6. Gas and dust production rates

### 6.1. Dust production rate: $Af\rho$ parameter

To characterize the dust abundance in the coma, the dust production proxy $Af\rho$ is widely used (A'Hearn et al., 1984), where $A$ is albedo of grains, $f$ is filling factor of grains within the field of view, and $\rho$ is the radius of the aperture at the comet. The $Af\rho$ parameter (in cm) is a measure of dust production, which is independent of unknown parameters, such as albedo and grain size. We calculated $Af\rho$ according to the equation (A'Hearn et al., 1984):

$$A(\lambda)f = \left(\frac{2r\Delta}{\rho}\right)^2 \frac{F_{com}(\lambda)}{F_{Sun}(\lambda)}. \tag{1}$$

Here, $F_{Sun}(\lambda)$ is the flux of the Sun at 1 au convolved with the filter transmission curve, $F_{com}(\lambda)$ is the flux of the comet in continuum. The value of $Af\rho$ is inferred from the RC′ images within the circular aperture of 4000 km, where there is the highest concentration of dust. The continuum flux was also determined from the long-slit spectrum at wavelength λ6835 Å which corresponds to the central wavelength of the continuum filter RC of the ESA filter set. The emission contamination was subtracted prior computing $Af\rho$. Calculating the aperture correction, we took into account the ratio between the slit areas and the circular aperture as well as the observed brightness distribution along the slit. The same aperture size was used for calculation of $Af\rho$ from the spectral observations. The errors in the $Af\rho$ parameter were calculated taking into account uncertainty of the comet flux introduced above. For this, spatial profile within the range free from emissions was used for estimation. Then it was recalculated to the standard wavelength, for which the $Af\rho$ value was determined in accordance with Eq. (1). The errors in $Af\rho$ were estimated from the spectrum using the expression:

$$\frac{\Delta(Af\rho)}{Af\rho} = \sqrt{\left[\frac{\Delta F(\rho)}{F(\rho)}\right]^2 + 2\left[\frac{\Delta f(\lambda)}{f(\lambda)}\right]^2}, \quad (2)$$

where $\Delta F(\rho)$ is the error in the total flux $F(\rho)$ along the slit height of 5000 km, $\Delta f(\lambda)$ is the error in the spectral distribution of the flux for the continuum $f(\lambda)$. The obtained results are presented in Table 4.

**Table 3**
The model parameters used to determine the gas production rates.

| Molecule | g-factor (watts/mol) | $l_p$ ($10^4$ km) | $l_d$ ($10^4$ km) | Power-law index $r^{-n}$ | Lifetime (sec) |
|---|---|---|---|---|---|
| CN(0–0) | $4.4 \times 10^{-20}$ | 2.19[a] | 30.00[a] | 2 | 339,940.0 |
| $C_2(\Delta\nu = 0)$ | $4.5 \times 10^{-20}$ | 1.60[a] | 11.00[a] | 2 | 124,645.0 |
| $C_3$ | $1.0 \times 10^{-19}$ | 0.10[a] | 6.00[a] | 2 | 67,988.1 |
| $NH_2$ (0,7,0) | $1.85 \times 10^{-21}$ | 0.41[b] | 6.20[b] | 1.55 | 70,254.4 |

[a] From Schleicher et al. (1987).
[b] From Cochran et al. (1992).

**Table 4**
Gas and dust production rates in comet 2P/Encke from spectral and photometric data on January 23, 2017.

| | $Q$ ($10^{25}$ mol/s) | | | | log $Q$ ratio | | | $Af\rho$ (cm) |
|---|---|---|---|---|---|---|---|---|
| | CN | $C_2$ | $C_3$ | $NH_2$ | $C_2$/CN | $C_3$/CN | $NH_2$/CN | RC |
| From spectroscopy | 3.19 | 2.66 | 0.17 | 0.61 | −0.08 | −1.00 | −0.72 | 92 ± 26 |
| From photometry | – | 3.87 | – | 1.04 | – | – | – | 57 ± 12 |

*6.2. Gas production rates*

In order to compute the production rate of gas molecules, we used the Haser model (Haser, 1957). The values of gas production rates are strongly dependent on the model parameters. We used the model parameters given in Table 3. There are the fluorescence efficiency (g-factor), parent and daughter scale lengths ($l_p$ and $l_d$, respectively), molecule lifetime, and the power-low index $n$ in the dependency on heliocentric distance ($r^{-n}$). The values of the parameters were taken from Langland-Shula and Smith (2011), except the parent and daughter length scales for $NH_2$ molecules which were taken from Cochran et al. (1992). The fluorescence efficiency $g$ for the CN(0–0) band depends on the heliocentric distance as well as the heliocentric velocity of the comet. Tatum (1984) provided the g-factor values for the CN molecule, taking into account the Swings resonance-fluorescence effect. More recently, Schleicher (2010) tabulated the g-factor values for various heliocentric distances and velocities which we used in our calculations. For the gas outflow velocity, we adopted the value 1 km/s used in most of such calculations (e.g., Langland-Shula and Smith, 2011). Since the Haser formula is valid for the circular aperture (15″ aperture size or 1.45 × $10^4$ km at the distance of the comet), and the spectrograph slit is not a circular one, for the correct application of the formula we included into the measured data a factor equal to the ratio of areas of the aperture used and the spectrograph slit within that aperture. Considering all these parameters, we calculated the molecular production rates of CN, $C_2$, $C_3$, and $NH_2$ in comet Encke. The results are presented in Table 4. There are also the ratios of the production rates of $C_2$, $C_3$, and $NH_2$ with respect to $Q$(CN). The log[$C_2$/CN] and log[$C_3$/CN] ratios are in a very good agreement with the data by A'Hearn et al. (1995) at $r$ = 0.84 au and Fink (2009) at $r$ = 0.55 au.

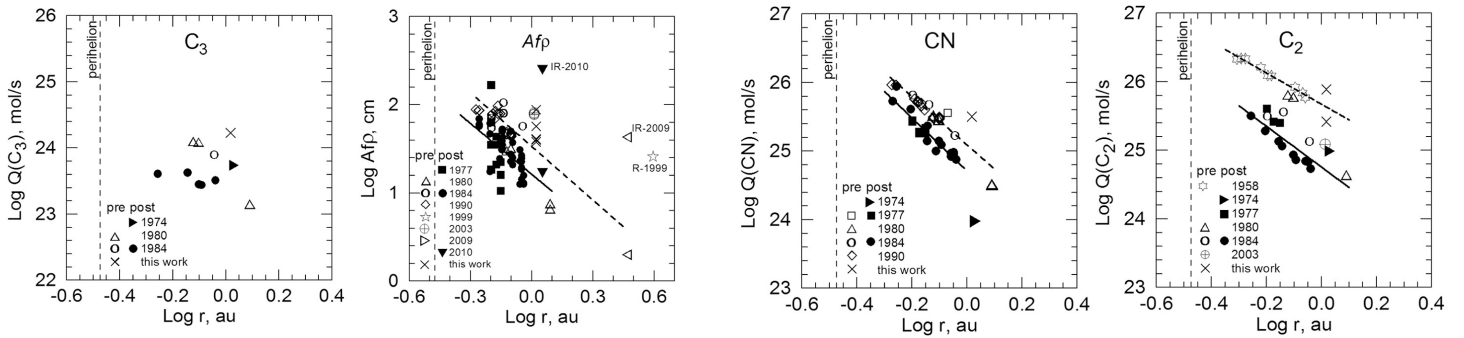

**Fig. 7.** Log of production rates of CN, $C_2$, and $C_3$ molecules and $Af\rho$ (as a measure of dust production), are plotted as a function of the log of heliocentric distance. All available data (Newburn and Johnson, 1978; A'Hearn et al., 1979; A'Hearn et al., 1983; A'Hearn et al., 1985; Osip et al., 1992; A'Hearn et al., 1995; Lowry et al., 2003; Reach et al., 2013) for the 1958, 1974, 1977, 1980, 1984, 1990, 1999, 2003, 2009, and 2017 (this work) apparitions of comet 2P/Encke are included for analysis. Open symbols indi-cate pre-perihelion observations, and filled symbols are observations taken after perihelion. Oblique crosses indicate data obtained in the present paper. Dashed and solid lines show a linear fit separately by least squares to the pre- and post-perihelion data, respectively. All species, except $C_3$, for which there are few observations, are well fitted by straight lines.

**Table 5**
Average values of the exponent $n$ in the gas (CN, $C_2$) and dust ($Af\rho$) production rates in comet 2P/Encke.

| $r$ (au) | Parameter $n$ in $Q \propto r^{-n}$ and $Af\rho \propto r^{-n}$ | | |
|---|---|---|---|
| | CN | $C_2$ | $Af\rho$ |
| Pre-perihelion 0.53–2.93 | 3.52 ± 0.29 (27)[a] | 2.28 ± 0.12 (9) | 2.03 ± 0.27 (38) |
| Post-perihelion 0.54–1.13 | 3.86 ± 0.31 (21) | 2.98 ± 0.64 (13) | 1.92 ± 0.33 (45) |

[a] Figures in brackets are the amount of data points.

### 6.3. $NH_2$ production rate

Comet Encke was also observed through the narrow-band filter $NH_2$ (λ6615/34 Å) and mid-band filter SED500 (λ5019/246 Å) that allowed us to calculate the production rate $Q$ of molecules $NH_2$ and $C_2$ using the Haser model. We used the same parameters (see Table 3) for both spectral and photometric data. The aperture size was $1.45 \times 10^4$ km. From photometric observations $Q(NH_2) = 1.04 \times 10^{25}$ mol/s and from spectral observations $Q(NH_2) = 0.61 \times 10^{25}$ mol/s at the heliocentric distance of 1.05 au. These results are presented in Table 4. Estimation of the gas production derived from the spectrum may differ from that obtained in the circular aperture due to the fact that spectrograph slit captures part of the fan. Somewhat larger value of $Q(NH_2)$, equal to $2.0 \times 10^{25}$ mol/s, was obtained by Fink and Hicks (1996) for comet Encke at the distance $r = 0.57$ au. For comparison, the value $Q(NH_2)$ for 9P/Tempel 1 was on average $(7.2 \div 9.6) \times 10^{24}$ mol/s (Weiler et al., 2007) at $r = 1.5$ au, for 81P/Wild 2, it was $(3.2 \div 4.24) \times 10^{25}$ mol/s at $r = (1.8 \div 1.6)$ au (Fink et al., 1999), and for 67P/Churuymov-Gerasimenko (hereafter 67P/C-G), it was $3.54 \times 10^{25}$ mol/s at $r = 1.6$ au (Ivanova et al., 2017b).

The $NH_2$ production rate was used to estimate $NH_3$ abundance in comet Encke (Tegler and Wyckoff, 1989). According to the photolytic reaction for ammonia, i.e. $NH_3 + h\nu = NH_2 + H$, the primary product of photodissociation of $NH_3$ is $NH_2$. Currently, there are no known "alternative" $NH_2$ parents with large production volumes. Therefore, the abundance of $NH_2$ is essentially a direct measurement of the abundance of $NH_3$ in comets. Kawakita and Watanabe (1998) showed that $Q(NH_3) = 1.05 \times Q(NH_2)$, according to Magee-Sauer et al. (1989). Based on this ratio, we found $Q(NH_3)$ for comet Encke equal to $1.09 \times 10^{25}$ mol/s from photometric and $0.64 \times 10^{25}$ mol/s from spectroscopic observations. Hatchell et al. (2005) derived the following upper limits of the $NH_3$ production rate for 5 comets from dynamically different groups: $<1.9 \times 10^{26}$ mol/s for C/2001 A2 (LINEAR); $2.7 \times 10^{26}$ mol/s for C/2001 Q4 (NEAT), $<2.3 \times 10^{27}$ mol/s for C/2002 T7 (LINEAR), and $\leq 6.3 \times 10^{26}$ mol/s for comet 153P/Ikeya-Zhang. Thus, the production rate of molecules $NH_3$ in short-period comet Encke is significantly smaller than that in the long-period comets.

### 6.4. Variations of gas and dust production rates with heliocentric distance

Although there is a very large volume of observational data for comet Encke, larger than for almost any other comet, it turns out that there is not much data on the gas and dust production rates that allow studying their dependences on the heliocentric distance. Comet Encke exhibits a brightness asymmetry about perihelion (e.g., Kresák, 1965); it is typically two magnitudes brighter before the perihelion than at the corresponding time after the perihelion. Having new data, together with the available data obtained from 1958 to 2010 apparitions, we investigated the variations of the production rates of CN, $C_2$, $C_3$, and $Af\rho$ in comet Encke with the heliocentric distance. In Fig. 7, the open symbols represent the pre-perihelion data and the filled symbols represent the post-perihelion data. Different symbols represent different apparitions but we have not distinguished the data by different authors. We made power-law fits to the pre- and post-perihelion production rates (see Fig. 7). The heliocentric distance dependence of $Q(CN)$, $Q(C_2)$, and $Af\rho$ is well represented in the log–log plot by a straight line. Averaged values of the exponent $n$ in the relation $Q \propto r^{-n}$ for gas and dust ($Af\rho$) production rates are presented in Table 5. The slopes of the production rates of the $C_3$ and $NH_2$ molecules in the Encke coma cannot be determined due to lack of suitable data.

In Fig. 7, we collected all available data on the production rates of CN, $C_2$, and $C_3$ molecules and $Af\rho$ for dust production in comet Encke and plotted as a function of the heliocentric distance. The $Af\rho$ values, as well as the CN value, resulted from our observations, are slightly higher than the data obtained earlier for this comet. Three points in the figure, derived from the IR observations by Reach et al. (2013) and in the R filter by Lowry et al. (2003), exhibit an anomalously large $Af\rho$ for the specified heliocentric distances and therefore we excluded them from further analysis. As can be seen in Fig. 7, comet Encke showed its typical asymmetry in the gas and dust production rates about perihelion, being less productive after perihelion, i.e. gas (and dust) release rates are higher pre-perihelion in comparison with post-perihelion at a given $r$. This difference in the production rates can be due to seasonal effects on nucleus.

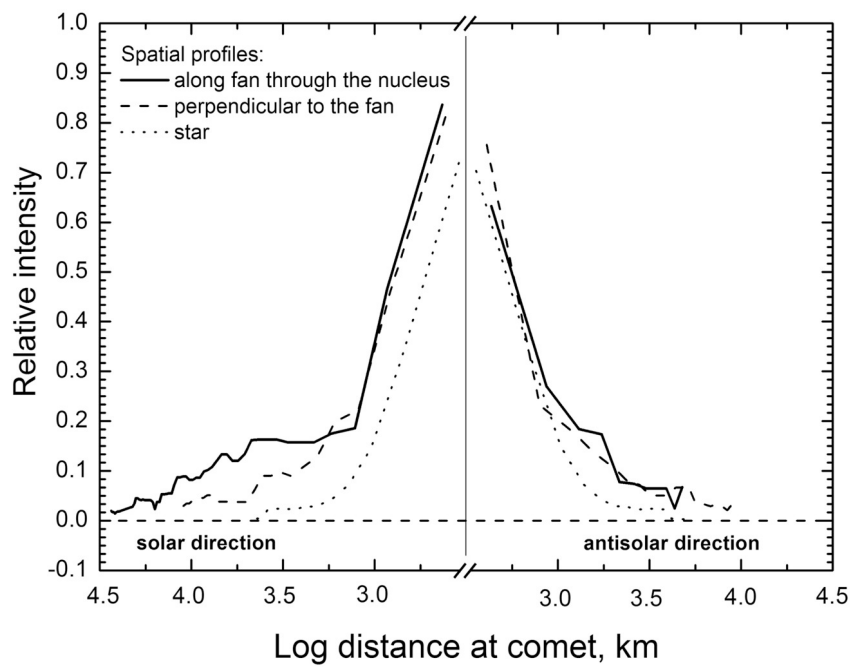

**Fig. 8.** Normalized surface brightness profiles of comet 2P/Encke and a field star. Profiles are taken across the central region of the comet along the fan (solid line), perpendicular to the fan direction (dashed line), and a field star (dotted line) taken from a single frame RC′. The negative values of distances display the direction toward the elongated fan in the solar hemisphere of the coma; the positive values are for the opposite direction.

We have found (Table 5) that CN shows steeper slopes than $C_2$, while $Af\rho$ varies with the distance from the Sun less steeply than the gaseous species. Such trends are similar to those detected for comet Encke by A'Hearn et al. (1995). According to A'Hearn et al., there are significant variations in the production of gas and dust, and respectively in the slope of log$Q$ (gas) and log$Af\rho$ with log$r$, from one comet to another, but on average, the variation of the production of gas is $Q \propto r^{-2.7}$ and of dust is $Af\rho \propto r^{-2.3}$. Given the large uncertainties in the determination of the power-law slopes, our results for comet Encke are insignificantly different from the results obtained by A'Hearn et al.

### Surface brightness profiles of dust

The surface brightness profiles of the isolated dust coma were obtained from the RC′ image (see Section 5.1), which is not contaminated by the gas. Fig. 8 demonstrates the radial cuts in the intensity images through the central region of the comet along the fan-antifan direction (solid line) and in the perpendicular direction (dashed line). The individual curves display the average coma flux in the 3 × 3 px size aperture with increasing projected cometocentric distance. The central region may be affected by the seeing corresponding to 1114 km. Therefore, we made the profile of a field star (dotted line) taken from a single frame RC′. The fluxes of the comet and star have been normalized to unity to compare how the profiles change with the distance from the nucleus and to search for any differences between the two profiles. The coma and fan are only slightly more extended than the instrumental point spread function, with little of the extension along the solar direction that would be expected for large slow-moving particles. The brightness drops in the circum-nucleus region is very steep, especially in the antisolar side. Comparison of the radial profiles shows that the coma is asymmetric and there is an excess in both "wings" of the comet profiles.

The surface brightness profiles of comet Encke presented in Fig. 8 show that the brightness distributions on the solar and antisolar sides are slightly different. The index $a$ in the dependence $I \propto \rho^a$ is on average −0.86 in the solar direction of the fan, whereas in the antisolar side $a$ = −1.03. In the direction perpendicular to the fan, $a$ is equal −0.95 and −1.01 in the solar and antisolar directions, respectively. Thus, the coma on the solar side was not in a steady state, while the close to a steady-state flow with a constant velocity was observed on the antisolar side.

### Contribution of the nucleus to the total brightness

It is possible that some, if not most, of the inner coma continuum observed in comet Encke is produced by the light reflected from the nucleus rather than from the grains. The comet was observed relatively close to the Sun and had an extended coma. Insufficient spatial resolution did not allow us to separate unambiguously the nucleus and coma signals, so we investigated possible contribution of the Encke's nucleus by convolving the images with the seeing profile to estimate the continuum that might be from a point source nucleus. For this, we compared the surface brightness profile of the comet with the profile of a field star, which is a proxy for the PSF (Fig. 8). To determine the contribution of the nucleus brightness to the total coma brightness, we used the method of fitting the model surface brightness to the observed images developed by Lamy et al. (2011). The observed surface brightness distribution in the near-nucleus region is a sum of the flux from the coma and cometary nucleus. The contribution of each of these two components was determined using seeing-convolved image of a model comet. We created images of model comets which possess the same image scale and PSF as the observed data, so that model profiles and real profiles can be directly compared. For representation of the real coma, we adopted a simplified approach, based on the isotropic model of the coma, in which its surface brightness $I_c$ varies according to a power law:

$$I_c = \frac{k_c}{\rho^a}, \qquad (3)$$

where $a$ is the power exponent of the brightness variation from the projected distance $\rho$ from the optocenter (see Section 7), $k_c$ is a scaling factor. Since the nucleus is not resolved, we will represent it as the Dirac delta function and its brightness in the adopted model can be written as:

$$I_n = k_n \delta(\rho), \qquad (4)$$

where $I_n$ is the contribution of the nucleus, i.e., the PSF scaled by the factor $k_n$. To take into account the real resolution (seeing) during the observations, we used the averaged star profile which corresponds to the point spread function of the telescope PSF. In this case, the model brightness distribution of the comet can be represented by a two-dimensional convolution:

$$I = (I_c + I_n) \otimes PSF, \qquad (5)$$

where $\otimes$ is the convolution operator.

**Table 6**
Model parameters for different power exponent $a$.

| Exponent $a$ | Coma $k_c$ | Nucleus $k_n$ | Error (obs – model) | Nucleus $f_0$ | Nucleus $f_{2000}$ |
|---|---|---|---|---|---|
| Surface brightness profiles along the fan | | | | | |
| −0.5 | 85 | 1115 | 11.8 | 0.76 | 0.47 |
| −0.7 | 115 | 937 | 12.0 | 0.63 | 0.40 |
| −0.9 | 150 | 667 | 12.6 | 0.45 | 0.31 |
| −1.0 | 169 | 482 | 13.0 | 0.33 | 0.24 |
| −1.1 | 191 | 244 | 13.5 | 0.17 | 0.13 |
| Surface brightness profiles perpendicular to the fan direction | | | | | |
| −0.7 | 92 | 1062 | 13.7 | 0.74 | 0.49 |
| −0.9 | 122 | 840 | 13.8 | 0.59 | 0.41 |
| −1.0 | 138 | 685 | 13.9 | 0.48 | 0.35 |
| −1.1 | 156 | 487 | 14.0 | 0.34 | 0.27 |
| −1.2 | 175 | 227 | 14.2 | 0.16 | 0.14 |

Since the nucleus center can be displaced relatively to the central pixel, i.e. the pixel having the largest signal, the model fitting was performed at sub-pixel resolution. The optocenter position of the comet was determined as an additional parameter of the convolution operator from comparison of the model calculations and observational profile. Since we have one-dimensional cuts, the displacement relative to the sub-pixel location of the nucleus was found with the help of the parameter $dr$. This parameter is the difference between the positions of the pixel center and brightness center of the comet in the pixel. The model profile depends on the position of the cometary nucleus in the central pixel. The value $dr = 0$ corresponds to the position of the nucleus strictly in the pixel center and provides symmetric model profile of the brightness distribution relative to the nucleus. For other values of $dr$, the symmetry of profile is broken.

The coefficients $k_c$ and $k_n$ were determined by the least squares method, minimizing the residuals between the model and the observed profiles. All coma profiles were fitted with power exponents ranging from $a = -0.5$ to $a = -1.3$. Larger values of $a$ lead to non-physical solutions. Table 6 represents the model results of determining the nucleus contribution for two profiles: along the fan-antifan direction and perpendicular to it. Here, the following designations are used: $a$ is the exponent in the model coma; $k_c$ and $k_n$ are coefficients of proportionality for coma and nucleus in the model; error (obs − model) is the root-mean-square error of the difference between the observed and model brightness distribution, $f_0$ is the contribution of nucleus radiation to central pixel intensity; $f_{2000}$ is the nucleus contribution to the integral intensity of the coma with radius 2000 km. The table shows that the nucleus contribution to the coma brightness strongly depends on the exponent $a$, because the shape of the central peak mainly depends on parameter $dr$ and is very similar to the response from a point source. The main difference between models with different parameters, but with the same $dr$, expected to be in the wings of the central peak. Since the central peak has large amplitude, its contribution to the total error of this model set of parameters should be quite large. Therefore, the total error changes weakly with a change of the model parameters. However, the set of model parameters, which provides a minimum error, is determined quite confidently. The best agreement between the model and observed profiles occurs at $a = -0.5$ for the fan direction and $a = -0.7$ for the perpendicular direction. In these cases, the nucleus contribution to the total coma brightness is 0.76 for the central pixel for the fan profile and 0.74 for the profile perpendicular to the fan direction. The contribution of the nucleus' flux to the integral intensity of the near-nucleus coma with radius 2000 km is on average 0.48. Thus, for the central pixel, the nucleus magnitude is $-2.5\log(0.75) = 0.3^m$ fainter than the total magnitude, whereas for the 2000 km near-nucleus area this is about $0.8^m$. In addition, independent determination of the dust coma contribution to the central pixel of two brightness profiles (along the fan and perpendicular to it) gives a fairly close estimates for different values of the power exponent *a*.

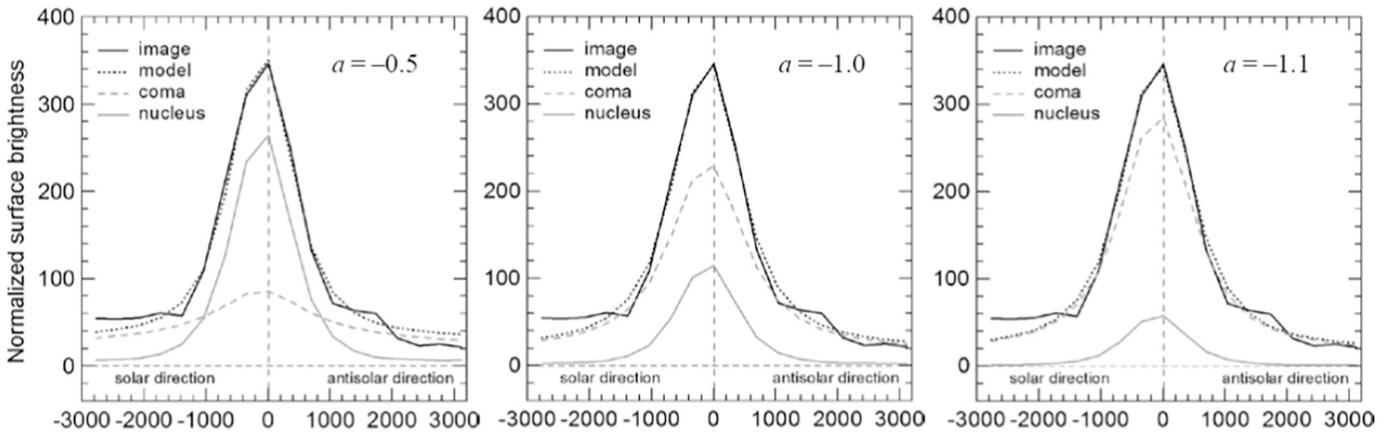

**Fig. 9.** Comparison of the model and observed profiles for different power exponents $a$. The continuum surface brightness profiles across the peak pixel of the comet 2P/Encke nucleus: top row are the profiles along the fan-antifan direction and bottom row are the profiles along perpendicular to the fan-antifan direction.

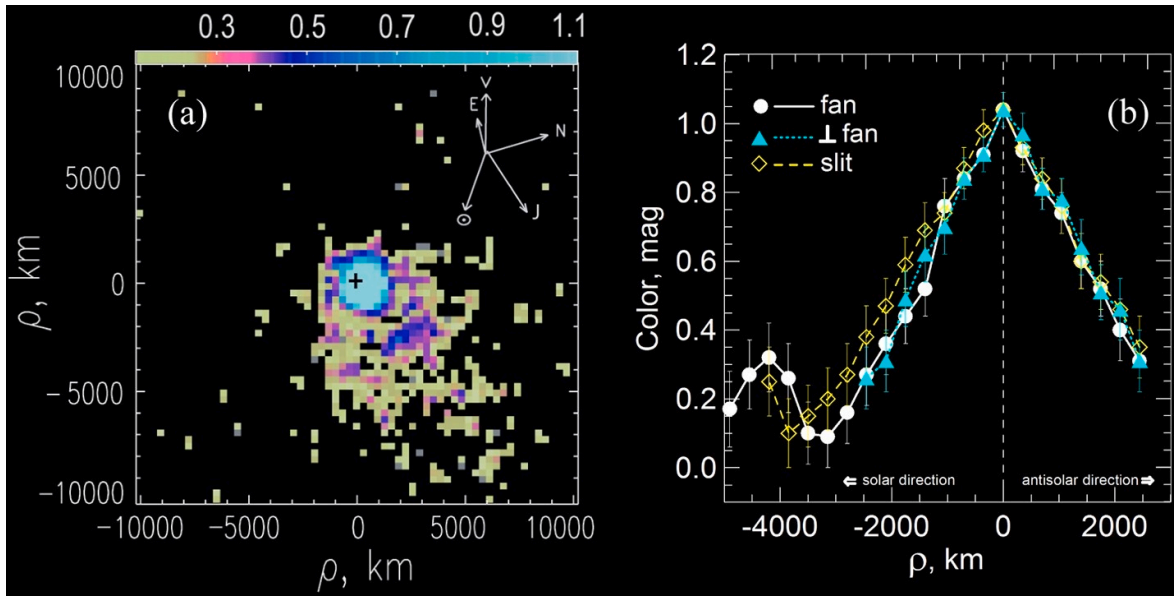

**Fig. 10.** Left panel (a) is the (BC–RC′) color map of comet 2P/Encke for January 23.75, 2017 with labels indicating the projected directions toward the Sun, North, East, jet (J), and the comet velocity vector V. The optocenter is marked with the white cross. Right panel (b) shows individual scans taken through the photometric center along the fan and perpendicular to it and along the direction of the spectrograph slit which coincides with the velocity vector.

The results of our modeling are also demonstrated in Fig. 9, where the brightness profiles across the central pixel along the fan-antifan direction and perpendicular to this direction are displayed. The model profiles along the fan and in the perpendicular direction give a fairly close value of the nucleus contribution with the mean value of 0.75. A choice in favor of model parameters $a = -0.5$ and $a = -0.7$ is also confirmed by the following. After taking into account the PSF function, the magnitude of the cometary nucleus is $18.8^m \pm 0.2^m$. This value was obtained from the RC-NH$_2$ image, in which the continuum contribution in the NH$_2$ image was taken into account. This contribution was equal to $0.59^m$. Thus, our result is in a very good agreement with the nucleus magnitude in the Cousins R filter $18.8^m \pm 0.5^m$. This value is obtained for our observations geometry ($r = 1.05$ au, $\Delta = 1.34$ au, $\alpha = 46.8°$), taking into account the phase coefficient of $\beta = 0.061$ mag/deg. and absolute magnitude of the nucleus $m_R(1,1,0) = 15.2^m \pm 0.5^m$ obtained by Fernández et al. (2000).

## 9. Color and normalized reflectivity

For calculations of the dust color in comet Encke, the images in the narrow-band cometary continuum filters BC and RC obtained on January 23, 2017 were used. In Section 5.1, we subtract the NH$_2$ image of the comet from the RC image and, thus, derive the RC′ image corrected for NH$_2$ contamination. The BC filter is relatively clean of gas emissions, although C$_2$ emission partially penetrates the BC passband. According to the spectrum of comet Encke (see Fig. 3), the emission/continuum ratio in the BC filter is more than twice smaller than in the RC band. Since we do not have C$_2$ emission image in the coma of comet Encke, we first plotted the (BC–RC′) color map of the comet (Fig. 10a), and after that considered the effect of gas contamination on the color profiles (Fig. 10b).

Using the central brightness peak in the coma defined from the isophotes with the accuracy 0.05 px, the flux calibrated images of the comet in both bands, BC and RC′, were carefully centered on the same position. Based on the derived images, we created a color map and analyzed color variations in the coma. We converted each pixel for summed images into the apparent magnitude and created the final (BC–RC′) color map by subtracting the two sets of images from each other. An average error in the magnitude measurements is 0.06$^m$. Fig. 10a shows a color map (BC–RC′) of comet Encke, while the right plot (b) displays the cuts across the color map of the comet along the jet and the direction perpendicular to the jet and along the direction of the spectrograph slit. As the distance from the nucleus increases, the color profiles show a strong color gradient: the color of the coma decreases sharply from about 1.1$^m$, but almost symmetrically around the nucleus, to a value of about 0.1$^m$ at the distance of ~3000 km along the fan direction. At distances approximately from 3000 km to 4000 km, a rather bright part of the shell (or cloud) is visible.

Based on the spectrum of comet Encke (see Section 3), we corrected the derived BC–RC′ profiles (Fig. 10b) for the contribution of the emission component to the BC passband, using the method for estimation of the dust/gas ratio described in Section 10 (Eqs. (10)–(20)). After taking into account this correction, the color-index for the total profiles of the dust coma and nucleus (denoted as BC′–RC′) increased in comparison with the profiles in Fig. 10b.

After that, we determined the color change of dust in the immediate vicinity of the cometary nucleus. For this, we used model calculations of intensity of the reflected light from the nucleus $I_n$ and dust coma $I_d$ as well as BC′–RC′ color profiles corrected for the emissions falling into the BC and RC passbands. To obtain the total color profiles for the dust and cometary nucleus, we used the following expression:

$$CI - CI_n = -2.5 \log \frac{I_d^{BC} + I_n^{BC}}{I_d^{RC} + I_n^{RC}} + 2.5 \log \frac{I_n^{BC}}{I_n^{RC}}, \quad (6)$$

where the BC and RC indices designate intensities of dust and nucleus in appropriate filters, and $CI$ is total color-index for the dust and cometary nucleus, while $CI_d$ and $CI_n$ are color-indexes of the dust and nucleus, respectively. The ratio of dust intensity to nucleus intensity can be estimated for the BC filter from Eq. (6):

$$\frac{I_d^{BC}}{I_n^{BC}} = 10^{-0.4(CI - CI_n)} \left( \frac{I_d^{RC}}{I_n^{RC}} + 1 \right) - 1. \quad (7)$$

The final expression for calculation of the dust color without the nucleus contribution is as follows:

$$CI_d = -2.5 \log \frac{I_d^{BC}}{I_d^{RC}} = -2.5 \log \frac{I_d^{BC}}{I_n^{BC}} \frac{I_n^{RC}}{I_d^{RC}} + CI_n. \quad (8)$$

The ratio $I_d^{RC}/I_n^{RC}$ is calculated from the model data (Table 6), $I_d^{BC}/I_n^{BC}$ from Eq. (7), and then we computed the color-index for the dust not affected by the nucleus contribution.

To evaluate the BC–RC color of the nucleus, we used the following data: the B–R color of the nucleus is $1.12^m \pm 0.08^m$ (Lowry and Weissman, 2007), the B–R color of the Sun is $1.005^m \pm 0.012^m$ (Ramirez et al., 2012), the color BC–RC of the solar analog is $1.276^m$ (Farnham et al., 2000). As a result of calculations, the BC–RC color of the nucleus of comet Encke is $1.39^m$. The radial profiles of the dust color without the nucleus contribution along the fan and perpendicularly to the fan direction are presented in Fig. 11. It can be seen that the maximum color-index of dust $1.43^m \pm 0.25^m$ is observed in the immediate vicinity of the nucleus and within the error limit nearly coincides with the color of the nucleus itself. At a distance of about 2500 km, the color-index sharply drops to $0.4^m$.

The normalized reflectivity gradient was calculated with the following expression (Jewitt, 2002):

$$S' = \frac{20}{\Delta \lambda} \frac{(10^{0.4\Delta m} - 1)}{(10^{0.4\Delta m} + 1)}, \quad (9)$$

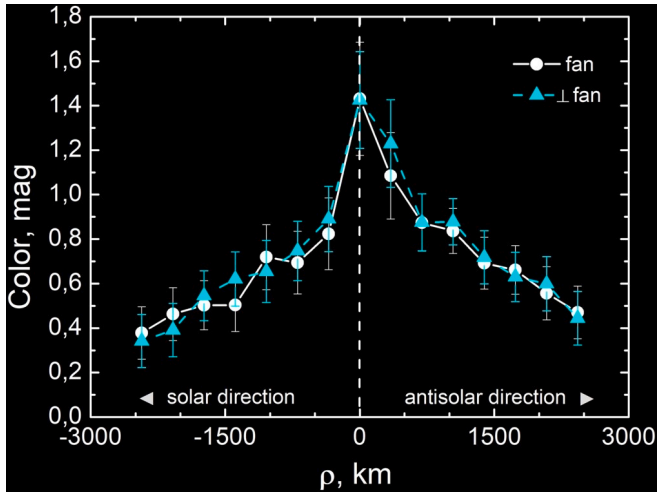

**Fig. 11.** The radial profiles of the dust color of comet Encke corrected for the emissions falling into the BC and RC passbands and for the nucleus contribution along the fan and perpendicularly to the fan direction.

where $\Delta m$ is the comet color minus the Sun color, $\Delta\lambda = \lambda_1 - \lambda_2$ is the difference in effective wavelengths. For comet Encke, we found reddening of the continuum with a gradient (9.4 ± 3.1) %/1000 Å and (8.5 ± 2.7) %/1000 Å from spectroscopy and photometry, respectively. Within the error margins, these values are consistent with the previous determinations of the reflectivity gradient of the Encke coma: (11 ± 2) %/1000 Å (Luu and Jewitt, 1990), (8.9 ± 1.6) %/1000 Å (Jewitt, 2004), (13 ± 7) %/1000 Å (Ishiguro et al., 2007), (11 ± 8) %/1000 Å (Boehnhardt et al., 2008). Note that the redness of the dust particles in the coma is larger than that for the nucleus of comet Encke, which was found to be (2.0 ± 4.5) %/1000 Å (on August 16, 1996) and (6.5 ± 3.5) %/1000 Å (on September 12, 1999) (Jewitt, 2002) and (2.5 ± 4.5) %/1000 Å (Meech et al., 2004), although Tubiana et al. (2015) found a higher reddening slope, (7.3 ± 0.2) %/1000 Å, in the wavelength range $\lambda$4200–7500 Å and (5.0 ± 0.2) %/1000 Å in the wavelength range $\lambda$4200–9000 Å for the bare nucleus of this comet at $r = 3.65$ au. This may suggest that the nucleus surface is depleted in organics in comparison with the dust particles, which contain some subsurface materials.

## 10. Spatial distribution of dust/gas ratio over the coma

Our aim is to investigate how gas and dust are distributed in the coma. In 2013, we observed a collimated outflow of substance in comet Encke that reached almost 50,000 km. Considering that radicals are formed from parent molecules and do not have a directed outflow and that the velocities of release of gas and dust are significantly different, such narrow gas jets (as can be seen in Fig. 4e) cannot be formed at a great distance from the nucleus. During the 2017 observations, we used narrowband continuum filters to avoid a contribution of gas emissions and could obtain distribution of pure dust over the coma; image is denoted as RC′ in Fig. 5d. In this image, the sunward fan-like coma is visible, which extends to at least ten thousand kilometers toward the Sun. Analyzing images taken with narrowband filters (Figs. 5 and 8), we have concluded (Sections 5 and 7) that dust is mainly concentrated in the near-nucleus region of comet Encke, but this does not mean that it is completely absent at larger distances, i.e. the dust may extend beyond a few thousand kilometers from the nucleus. The study of the composition (gas or dust) of structural features (fan/jets) in a coma is difficult because of the effects of projection. Therefore, we used the long-slit spectra to determine the dust-to-gas ratio as a function of distance and thus define the expansion of the dust coma. And below we present observational evidences for the presence of dust coma up to a few tens of thousands of kilometers.

As it is well known, spectra of comets consist of emissions caused by reemitting of solar photons by molecules (resonance fluorescence) and continuum caused by the scattering of sunlight by cometary dust particles. To separate these components, we used long-slit spectrum of comet Encke acquired almost simultaneously with imaging on January 23, 2017. As a result of processing the two-dimensional cometary spectrum, we derived the spectral flux density from the cometary coma for the emission component $i_e(\lambda)$ and the continuum $i_c(\lambda)$. Knowing the bandwidth of the used filter $p(\lambda)$, normalized to maximum transmission, we can estimate the contribution of emissions $I_{0e}$ and continuum $I_{0c}$ to the total radiation flux which is obtained from the central part of the cometary coma:

$$I_{0e} = \int p(\lambda) i_e(\lambda) d\lambda; \quad I_{0c} = \int p(\lambda) i_c(\lambda) d\lambda. \tag{10}$$

To determine the spatial distribution of the selected emissions and continuum, we extracted from the original spectra the intensity $i_e(\rho)$ of emission at the wavelength $\lambda_e$ with a maximum band intensity (to reduce the errors) and the continuum nearest to this emission $i_c(\rho)$ at the wavelength $\lambda_c$. Here, $\rho$ is the projected distance from the comet nucleus. The spatial profile of the continuum was corrected for the reddening effect due to the difference in wavelengths for the spatial profiles of emission and continuum:

$$i'_c(\rho) = i_c(\rho) \frac{i_c(\lambda = \lambda_e)}{i_c(\lambda = \lambda_c)}. \tag{11}$$

These spatial profiles can have a constant component due to the background signal. To take it into account, the mean value of the signal at the edge of the profile was subtracted from the observed spatial profiles:

$$i''_e(\rho) = i_e(\rho) - \overline{i_e(\rho \geq \rho_{\max})}, \quad i''_c(\rho) = i'_c(\rho) - \overline{i'_c(\rho \geq \rho_{\max})}. \tag{12}$$

To obtain a pure emission profile, $i'''_e(\rho)$, the continuum profile $i''_c(\rho)$ was subtracted from the emission profile:

$$i'''_e(\rho) = i''_e(\rho) - i''_c(\rho) \tag{13}$$

To determine profiles in a broadband filter, in which a single emission band predominates, the spatial profiles of emission $I_e(\rho)$ and continuum $I_c(\rho)$ components in this filter can be calculated according to the following expressions:

$$I_e(\rho) = I_{0e} \frac{i'''_e(\rho)}{i'''_e(\rho = 0)}, \quad I_c(\rho) = I_{0c} \frac{i''_c(\rho)}{i''_c(\rho = 0)}. \tag{14}$$

If several (j) emission bands penetrate in the filter, then the spatial profiles of the emission component are calculated taking into account these bands:

$$I_e(\rho) = I_{0e} \sum_{j=1}^{k} \frac{i'''_{ej}(\rho) w_j}{i'''_{ej}(\rho = 0)}, \tag{15}$$

where $w_j$ is the contribution of the $j$ emission band to the total spatial profile.

$$w_j = \frac{i'''_{ej}(\rho = 0)}{\sum_{n=1}^{k} i'''_{en}(\rho = 0)}. \tag{16}$$

We calculated the dust/gas ratio in the broadband filters SED500, r-sdss, and V. The spatial profiles of the continuum $F_c(\rho)$ for these filters were calculated from the profiles $f_c(\rho, \lambda_c)$, which were determined from the spectrum for a selected wavelength $\lambda_c$ free from the emission influence. For this, the previously obtained continuum spectrum $F_c(\lambda)$ and the known passbands $p(\lambda)$ of the filters were used:

$$F_c(\rho) = \frac{f_c(\rho \lambda_c) \int p(\lambda) F_c(\lambda) d\lambda}{F_c(\lambda_c)}. \tag{17}$$

To obtain the total spatial emission profile, we used the spatial profile $f_e + f_c(\rho, \lambda_c)$ for a given emission. The index e + c indicates that the profile contains the emission component and continuum. The emission profiles without continuum were calculated as follows:

$$f_e(\rho, \lambda_e) = f_{e+c}(\rho, \lambda_e) - f_c(\rho, \lambda_c) \frac{F_c(\lambda_e)}{F_c(\lambda_c)}. \tag{18}$$

In the case when a single emission falls into the filter passband, the spatial emission profile for this filter was determined as follows:

$$F_e(\rho) = \frac{f_e(\rho \lambda_e) \int p(\lambda) F_e(\lambda) d\lambda}{F_e(\lambda_e)}. \tag{19}$$

For the case of several emission bands, the spatial emission profile was calculated using the following formula:

$$F_e(\rho) = \frac{\int p(\lambda) F_e(\lambda) d\lambda}{\sum_i F_e(\lambda_e^i)^2 p(\lambda_e^i)} \sum_i F_e(\lambda_e^i) p(\lambda_e^i) f_e^i(\rho \lambda_e^i) \tag{20}$$

And finally, the spatial profile of the relation $k(\rho)$ of the continuum to the emission component for the selected filter is $k(\rho) = F_c(\rho)/F_e(\rho)$.

An examination of the transmission curves of the filters and the spectrum of comet Encke (see Fig. 1) shows that the main contributors to the SED500 and V spectral bands are the light scattered by dust particles and the resonance fluorescence by $C_2(0,0)$ molecules, whereas the r-sdss filter transmits continuum and $C_2$ plus $NH_2$ emission radiation. Spatial profiles of the $C_2(0,0)$ emission were produced along the slit by summing over the spectral extent of the $C_2(0,0)$ band (Fig. 12a, c), while profiles of the total radiation from ($C_2$ + $NH_2$) were defined for the r-sdss filter (Fig. 12e). The continuum spatial profiles were also generated for each filter with the emissions subtracted out. To do the subtraction correctly, we took into account the different light transmission of the filters and the different continuum levels. The sky levels for the emission and the continuum profiles were estimated and subtracted. As a result, we separated the clean spatial profiles of the $C_2$ and ($C_2$ + $NH_2$) emissions without contribution from the dust component and continuum without emissions (Fig. 12, left). After that we obtained the spatial distribution of the continuum/emission ratio (Fig. 12, right). To avoid cluttering, we show the error bars only through a certain number of data point. As one can see, our procedure yields a clean emission profile and the dust/gas ratio with reasonable accuracy, when the $C_2$ emission is well isolated. In the case of the r-sdss filter, the dust/gas ratio is obtained with much lower accuracy due to inaccurate accounting emission radiation of $C_2$ and $NH_2$.

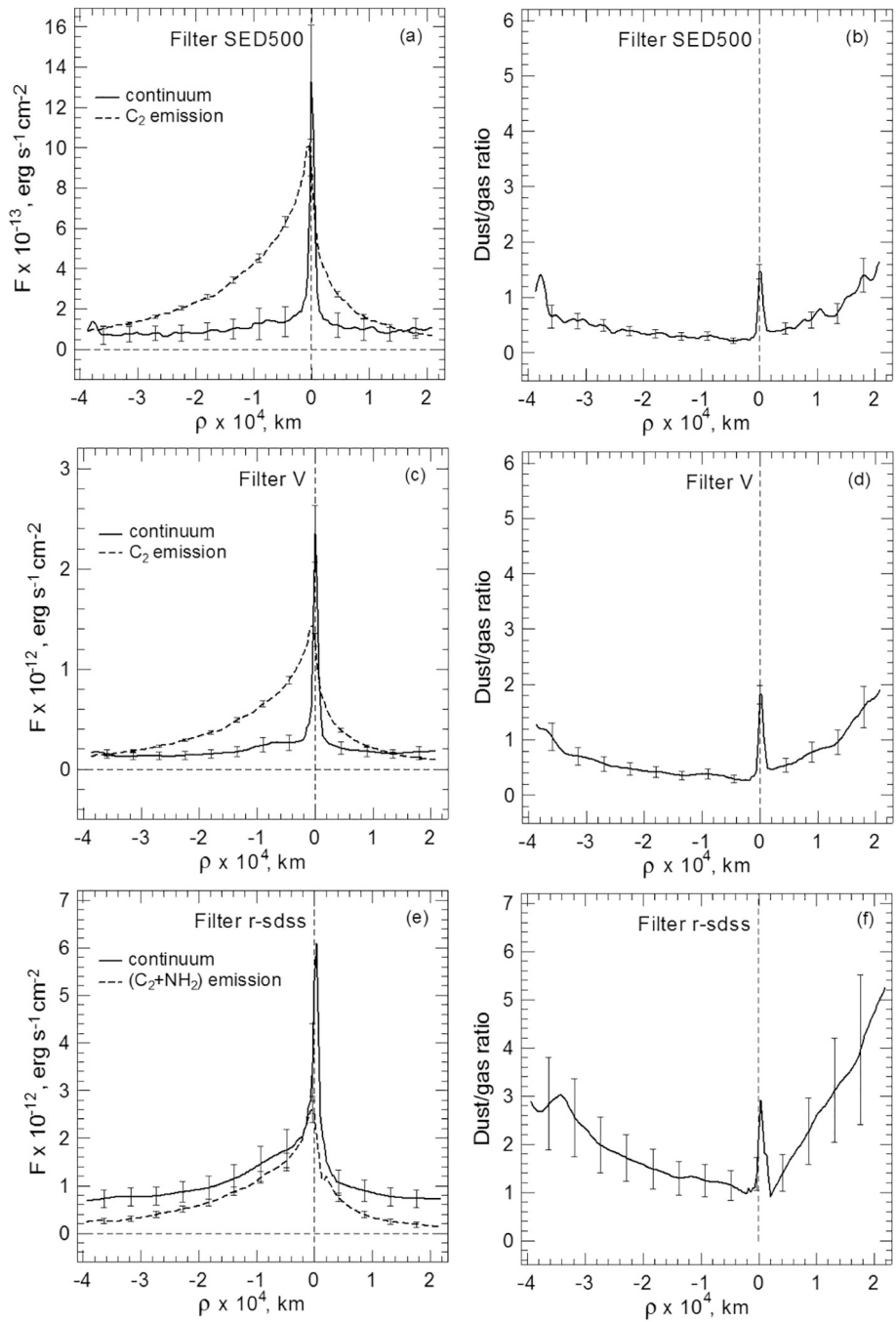

**Fig. 12.** Nucleus-centered spatial profiles of the $C_2$ (a, c) and ($C_2$ + $NH_2$) (e) emissions and corresponding continuum along the spectrograph slit. Sunward (negative distance) and antisunward (positive distance) spatial profiles are plotted as a function of the distance from the nucleus. On the right, the continuum-to-emission ratio is displayed (b, d, f). The error bars are shown through a certain number of data point.

**Table 7**
Power exponent $a$ for different cometocentric distances.

| Solar direction | | Antisolar direction | |
| --- | --- | --- | --- |
| log $\rho$ | $a$ | log $\rho$ | $a$ |
| 3.38–3.72 | 0.52 ± 0.08 | 3.38–3.86 | 0.31 ± 0.01 |
| 3.72–4.33 | 1.08 ± 0.02 | 4.13–4.44 | 1.18 ± 0.02 |

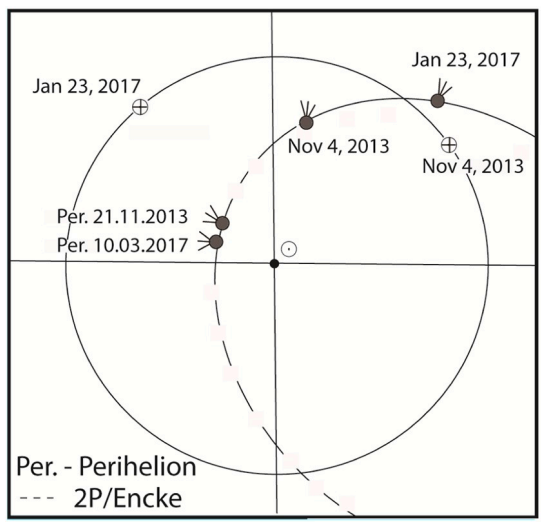

**Fig. 13.** Schematic view of the orbit of comet 2P/Encke along with the Earth's orbit during the observations in the 2013 and 2017 apparitions. The positions of Encke at the time of observations and the moment of the perihelion passage are marked by the comet symbols, and the Earth at the moments of observations is indicated by the Earth symbol.

The spatial profiles of the gas emission $C_2$ and of the continuum show that Encke's coma is asymmetric (Fig. 12). They also show that the dust is mainly concentrated in the near-nucleus coma, although it extends almost throughout the entire coma. The continuum profiles drop precipitously to approximately 2500 km and then very slowly expand to the distances exceeding $10^4$ km. Since *FWHM* of seeing is 1114 km, we believe that there is no influence of the seeing on strong drop of continuum at a distance of 2500 km from the nucleus. In Table 7, we give the power exponent $a$ for spatial profiles of the $C_2$ emission for different cometocentric distances $\rho$. One can see that for the $C_2$ spatial profile, the exponent $a$ is significantly smaller than 1 at small distances from the nucleus, and at large distances >1 for both directions. The characteristic feature is that the antisunward profiles are much steeper than those on the sunward side at distances from the nucleus larger than 10,000 km. Such behave of $C_2$ profiles may suggest that the gas production was not steady and indicate production of this radical from parent molecules and/or grains near the nucleus resulted from their destruction by photodissociation or photoionization at larger distances from the nucleus (Combi et al., 1997).

In addition to obtained spatial profiles, we calculated the averaged dust/gas ratio for each filter which we used during the observations. For this, we determined the fluxes from emission and continuum components in a given filter using the central part of the spectrum of comet Encke. As a result, the following values of the continuum/emission ratio are obtained: 3.14 in the BC filter; 1.56 in RC; 1.48 in SED500; 1.79 in V; 2.91 in r_sdss. These values are close to those obtained from spatial profiles in the near-nucleus areas (Fig. 12, right).

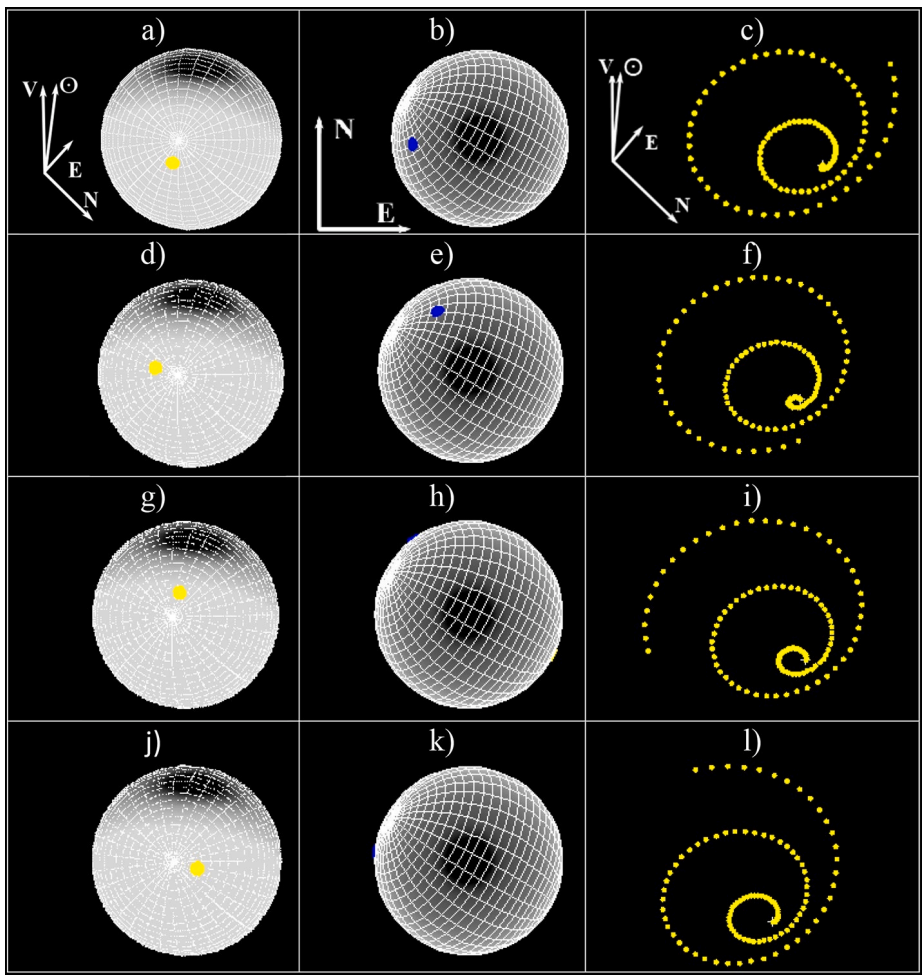

**Fig. 14.** Schematic representation of the nucleus of comet 2P/Encke together with the rotational axis as viewed from the Earth (a, d, g, j) and from the Sun (b, e, h, k) and the model jets obtained during various phases of the nucleus rotation (c, f, i, l) at the time of November 4.08, 2013. A yellow dot shows the active region located at the north hemisphere at the cometocentric latitude +55° (Source 1), whereas the active region at the south hemisphere at latitude −75° (Source 2) is depicted as a blue dot. Bright spots are the areas illuminated by the Sun at the considered moment. Panels (a, b, c) correspond to zero rotational phase; (d, e, f) correspond to the orientation after 0.25 of the rotation period; (g, h, i) are for 0.50 of the rotation period; and (j, k, l) are for 0.75 of the rotation period. The arrows indicate the direction to the Sun (☉), North (N), East (E), and velocity vector of the comet in projection on the sky (V). (For interpretation of the references to color in this figure legend, the reader is referred to the web version of this article.)

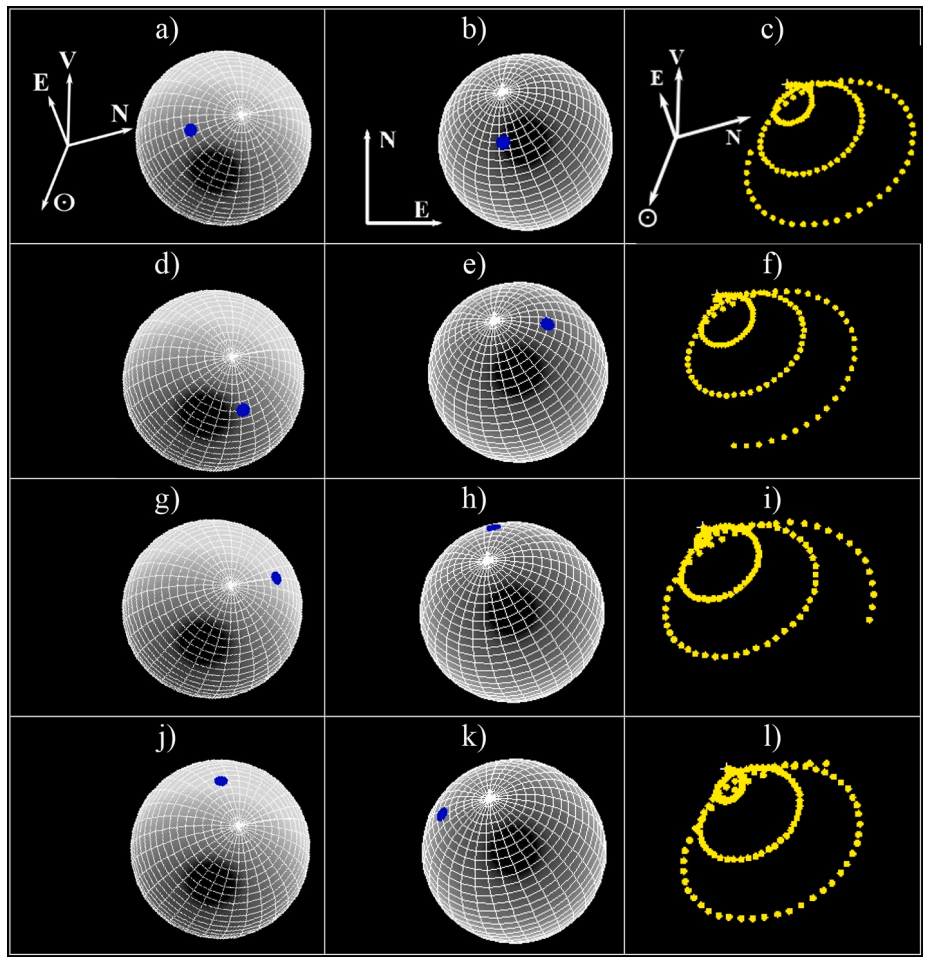

**Fig. 15.** Schematic representation of the nucleus of comet 2P/Encke together with the rotational axis as viewed from the Earth (a, d, g, j) and from the Sun (b, e, h, k) and the model jets obtained during various phases of the nucleus rotation (c, f, i, l) at the time of January 23.81, 2017. The active region Source 2 at the cometocentric latitude −75° cannot be observed form the Earth and do not appear on the hemisphere illuminated by the Sun. The other notations are the same as in Fig. 14.

**Modeling of dust emission geometry**

According to the Sekanina's model (Sekanina, 1988a), a change of the Encke surface exposed to the Sun (seasonal change) occurs around the perihelion due to the comet nucleus spin axis orientation. This lasts only a few weeks, due to the small heliocentric distance and the large orbit eccentricity. During the rest of the orbital period, lasting some years, the comet surface was almost at the same geometric conditions with respect to the Sun (Epifani et al., 2001).

Our observations of comet Encke were obtained 17 days before the perihelion in 2013 and 45 days before the perihelion passage in 2017. Fig. 13 shows the schematic view of the orbit of comet Encke and the Earth's orbit during our observations in the 2013 and 2017 apparitions. According to Sekanina (1991), the south pole of the nucleus is seen from the Sun from 8 days before perihelion to 60 days after, whereas the Sun passes in the nucleus equatorial plane ~10 days before perihelion and ~60 days after perihelion (Festou and Barale, 2000). Thus, our observations in 2013 were obtained with an illuminated equatorial region, whereas the 2017 data were taken when the nucleus north pole was illuminated by the Sun.

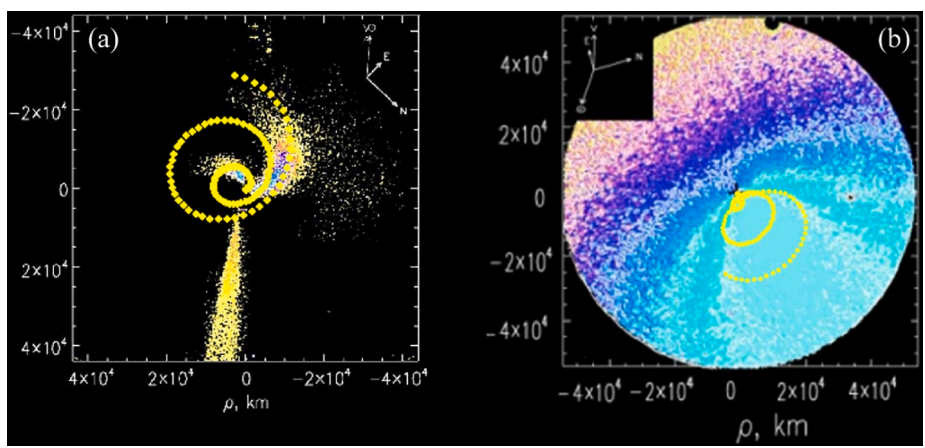

**Fig. 16.** Comparison of the jets observed on November 4.08, 2013 (a) and on January 23.78, 2017 (b) with the model jets (blue dots) from a discrete emission source on the nucleus. (For interpretation of the references to color in this figure legend, the reader is referred to the web version of this article.)

To better understand how active areas release matter from the nucleus and how the fan and jets can be formed, we need to model geometry of the viewing conditions and the rotation state of the nucleus. According to Sekanina (1987), asymmetric outgassing is observed in comets due to collimated ejection from discrete active areas on the sunlit side of rotating cometary nuclei. These asymmetric structures can be spiral jets, semicircular halos, straight plumes, sharp spikes, snail-shell-like features, and fans with or without jets. Analyzing the apparitions of Encke from 1924 to 1984, Sekanina, 1988b determined the orientation of Encke's sunward-facing fan, which is seen at almost every apparition, by identifying two discrete emission sources in the north and south hemispheres with the cometocentric latitudes of $+55°$ and $-75°$, respectively, and by coordinates of the north pole of the comet spin axis, right ascension $RA \approx 200°$, declination $Dec \approx +20°$ (apparitions in 1957–1984) and $RA \approx 205°$, $Dec \approx +10°$ (apparitions in 1924–1951). These results were confirmed by Festou and Barale (2000). However, later (2003 apparition) Woodney et al. (2007) and Ihalawela et al. (2011) did not confirm the presence of two active areas on the surface of the comet's nucleus and instead suggested the possibility of a single active area which forms single-cone jet with an opening angle of $90°$.

For analysis of the jet structure of comet Encke, we propose a simple geometrical model, which accounts for a rotation period of the cometary nucleus, the orientation of the comet spin axis, the cometocentric latitude of the source, and the relative position of the Sun, Earth, and the comet. We assumed that the nucleus is spherical, isolated active region releases gas and dust only under the solar irradiation, and the outflow of matter occurs radially from the nucleus with a constant velocity; dispersion of the velocities of particles ejected from the active region and their acceleration under the solar radiation were not considered. Such an approach is valid at small distances from the cometary nucleus. During the simulations, model jets, which reproduce the middle of the visible jet, are projected onto the plane of the sky. Sequence of images obtained with this procedure shows the motion of the ejected matter on the sky plane and jet evolution in time (Figs. 14 and 15).

We modeled the continuous ejection of dust particles from two discrete active regions, Source 1 and Source 2, to search for conditions that can reproduce the observed jets and fan and to explain their emerging. The model jets were compared with the structures observed close to the cometary nucleus at digitally processed images (Fig. 16). For the observations on November 4.08, 2013, the best fit between the observed coma morphology and model image was achieved under the following parameters: the rotational period of the nucleus was 11.1 h, the coordinates of the north rotation pole were $RA = 200° \pm 5°$ and $Dec = +5° \pm 3°$, escaping velocity of the dust particles from discrete region on the nucleus was $0.37 \pm 0.05$ km/s, and the emission source was located at the cometocentric latitude $+55°$, i.e., it was Source 1 (Fig. 14). The active Source 2 did not affect the jets because all the time it was in the shade whereas the active Source 1 was located on the illuminated hemisphere (see Fig. 14). However, in fact Source 1 was balancing on the edge between shadow and light during a significant part of the rotational period. This led to a small intensity of the ejections in the antisolar direction. Furthermore, it is possible that the sunlight was not reaching the area located on the terminator due to the topography of the nucleus.

As a result of the simulations, we revealed an interesting structural feature of the coma in the observations of the 2013 apparition. It turned out that two visible jets were formed by a single active region, Source 1 (Fig. 16a). The brightest jet was formed by two ejections at two sequential rotations when the active region was oriented toward the solar direction. In this case, the ejection from the heated region was stronger. The less bright jet was formed when the Source 1 entered the hemisphere illuminated by the Sun. This region just started to warm up and the release of gas and dust was at a lower level, therefore, the second jet had a lower brightness and was less extended.

The trajectory of a dust particle depends on the particle ejection velocity and β, which is the ratio of the solar radiation pressure $F_r$ to the solar gravity $F_g$. To estimate the effect of the solar radiation pressure on the velocity, the acceleration of the ejected particles by the solar radiation pressure, $a$, was computed following a relation $a = (\beta g_{au})/r^2$, where $g_{au}$ is the acceleration by the solar gravity at 1 au, $r$ is heliocentric distance. The particle velocity depends on the direction of the ejection and time after the ejection. The maximum decrease in velocity by solar radiation pressure is expected to be in the sunward direction. Sekanina and Larson (1984) gave an estimation $\beta \leq 0.6$ for cometary particles. Assuming $\beta = 0.3$ for our estimates, for a given heliocentric distance of the comet and jets geometry, the maximum decrease in the velocity should be 0.08 km/s for the outer parts of the jets. Since the particle velocity in the jet is estimated according to the shape of the whole jet, the actual decrease in particle ejection velocity is smaller. Thus, the value $0.37 \pm 0.05$ km/s is the lower estimate of the real velocity of the substance outflow from the cometary nucleus. This value is close to the values derived by other authors for comet Encke. According to Sarugaku et al. (2015), when comet Encke was at the heliocentric distance $r = 2.29$–$2.53$ au (apparition of 2004), the ejection velocity of the dust particles was typically 0.4 km/s. Lisse et al. (2004) estimated a typical velocity at $r = 1.04$–$1.12$ au equal to $0.5 \pm 0.1$ km/s. In both papers, a strong dependence of ejection rate on particle size was noted: for large particles, the ejection rate was lower. In our case, the brightest structure had a rather long formation time, about 3 nucleus rotations (~30 h). During this time, the fine dust fraction significantly changed its velocity relative to the nucleus due to the acceleration by solar radiation. The courser dust fraction kept the direction and velocity practically unchanged. Therefore, we should expect a slightly lower velocity of the dust particles than it is obtained from our model.

The best correspondence between the model jets and the coma structure observed on January 23, 2017 was obtained for the coordinates of north rotation pole $RA = 200° \pm 6°$ and $Dec = +25° \pm 5°$. It was found that the velocity of the ejected material from the active region for a given position of the comet does not affect the coma appearance. As in the apparition of 2013, only the Source 1 was active (see Figs. 14 and 15). However, in contrast to the November 4, 2013 observation, when the axis of rotation was directed almost along the line of sight, on January 23.78, 2017, the rotation axis was located close to the sky plane. Therefore the outflow of the dust from the Source 1 was within a jet which was seen as a cone in the projection onto the plane of the sky. Axis of this cone was significantly deviated from the sunward direction and shifted toward the projection of the rotation axis on the sky plane (Fig. 16b). This effect was responsible for the unusual shape of the cometary coma in January of 2017. In his paper, Farnham (2009) displayed the CN image of comet Encke (see Fig. 13 in that paper) obtained by Woodney et al. (2007) during the close approach of the comet in 2003. The spatial scale was sufficient to reveal a corkscrew morphology in the sunward fan and to show that the fan was indeed a cone of material centered on the spin axis.

## Discussion

Comet 2P/Encke has a number of features that attract attention of researchers. First, comet Encke is a highly evolved comet. Second, according to the observations in the visible wavelengths, it has an extremely low content of dust, which is concentrated in the near-nucleus region of the coma, making Encke one of the gassiest comets. And third, it has a sunward-oriented fan which has been observed near perihelion in almost all apparitions. Our observations aimed a better understanding of the features and processes in the coma of comet Encke.

*Morphology*

As can be seen from our observations on November 4, 2013, comet Encke had extended rounded coma with a plasma tail ($PA = 128.9°$) and two jet-like structures ($PA = 42.8°$ and $173.5°$) which were highly curved. The rotation axis of the nucleus was directed almost along the line of sight. Due to the solar wind, the gas tail should be straight and directed in the opposite direction from the Sun, that we see in Fig. 4. The revealed jets are not straight and are formed rather by grains than by gas molecules. The existence of the jets indicates that the nucleus has isolated active areas. In Section 11, we showed that the same active source formed the observed two jets. According to Lisse et al. (2005), the shape and appearance of jets are dependent on a number of factors: the size and location of the source: the rotation state of the nucleus, the motion of the comet in its orbit, and changes in the viewing geometry of the comet. Analysis of these factors shows that there are two factors which may produce the curvature of the observed jets in comet Encke: nucleus rotation and separation of particles by mass due to radiation pressure. The effect of nucleus rotation should dominate at small distances from the nucleus. The cometary dust tail is composed of particles of different sizes swept back by solar radiation pressure. It is possible that the brightness of the dust tail was below detection threshold, although other observers did not detect a dust tail in this apparition too (e.g., Vervack et al., 2014). Since the number density of dust particles in jets is substantially higher than in the surrounding coma, we could reveal them by digital filtering as shown in Fig. 4e.

In contrast to the 2013 apparition, on January 23, 2017, the images of the comet displayed a prominent sunward fan ($PA = 290°$), extending to ~30,000 km from the nucleus. The rotation axis of the nucleus was located close to the sky plane. The same fan was observed by Kwon et al. (2018) on February 19–21, 2017; they called it "dust cloud". Neither Kwon et al. nor we detected the dust tail, that might be evidence of absence of a small dust population that could be extended along the antisunward direction under the solar radiation. Most likely, the fan consisted of a population of compact large grains ejected from an active region of comet Encke. According to Sekanina (1979), an active source is located on the north hemisphere at pre-perihelion times. Our modeling the outflow of substance from the sunlit hemisphere of the nucleus show that in both observational periods the outgassing from the active region was from the same source located at the cometocentric latitude $+55°$, but due to differences in the position of the nucleus rotation axis relatively to the Sun and the picture plane in November 2013 and January 2017, the different structure of the coma was observed.

*Gas-to-dust ratio*

As our images of comet Encke in narrow continuum filters show (Fig. 5), the dust was mainly concentrated in the near-nucleus region where dust/gas ratio was about 3 in the r-sdss filter (Fig. 12). However, a high concentration of dust around the nucleus does not mean that it is completely absent at larger distances from the nucleus. As it was shown in Section 10, the dust/gas ratio drops sharply to distances of 2000–3000 km and then gradually increases, i.e. the dust extended almost throughout the entire coma, out to ~40,000 km. Even though dust particles are altered due to loss of volatiles or/and scattering cross-section change after sublimation and fragmentation, their distribution should peak sharply at the cometary nucleus. At the same time, gas molecules observed in the visible are daughter molecules and close to the nucleus they have a shallower distribution than cometary dust particles. It is expected in gas-rich comets because the gas/dust ratio increases with distance from the nucleus. Such behavior of dust and gas we see in comet Encke (Fig. 12) that allows us to make a conclusion about noticeable contribution of the dust component to the observed luminosity of comet Encke. It is particularly seen when the observations are performed with the filters with a relatively wide transmission curve and on the periphery of the cometary coma. This means that the dust contribution must be taken into account interpreting observations of even "gas comets" such as comet Encke. This especially concerns the nature of the jets (gas or dust?) which were detected in the apparitions of 2013 and 2017. An argument in favor of the dust jets is our model calculations, which showed that the coordinates of the active region were the same for both apparitions and coincided with the Sekanina, 1988b data. The found coordinates of the north pole of the nucleus are also consistent with the data by Festou and Barale (2000).

Estimate of the relative contribution of light scattered from the nucleus to the total brightness of the near-nucleus coma show that about 75% of the flux of the scattered light in the central pixel was due to the nucleus, and the nucleus flux contributed 48% to the total intensity of the coma of radius 2000 km. This means that the nucleus was $0.3^m$ fainter than integral magnitude of the central pixel and $0.8^m$ fainter than the 2000 km area of the coma. We determined that the nucleus magnitude of comet Encke was equal to $18.8^m \pm 0.2^m$ that is close to estimation from the HST data (Fernández et al., 2000). According to Belton et al. (2005), the nucleus contributed only a fraction to the total light, usually >28% but rarely over 87% at heliocentric distances 3–4 au. Thus, though the nucleus region includes contamination from the coma, the low activity of the comet at the time of observation suggests that the nucleus itself could make a noticeable contribution to the coma profile within the central seeing disc. This effect is well seen in Fig. 12a, which shows a strong central spike, caused by the unresolved nucleus, rising above a fainter profile of $C_2$. A similar effect was observed in comet 49P/Arend-Rigaux by Jewitt and Meech (1985).

*Spatial profiles*

Model spatial profiles along and perpendicular to the observed fan showed slightly different values of the $a$ index in the dependence $I \propto \rho^a$. The power index $a$ is significantly smaller than 1 and is equal to $-0.5$ along the fan and $-0.7$ perpendicular to it in the near-nucleus region, while from the observed image, $a$ varied within the range $-0.86 \div -1.03$ for different directions, including the outer parts of the coma. Perhaps, some difference in the $a$ index indicates different physical properties of the dust in the active region and in other parts of the cometary nucleus. Similar results for comet Encke were obtained by Sekanina and Schuster (1978) on September 12, 1977: $a$ varied from $-0.64$ to $-0.86$. On October 9–10, 1980, a shallow power-law decline of surface brightness of comet Encke occurred with the slope $-0.5$ up to 7000–10,000 km (Djiorgovski and Spinrad, 1985). Such low values of power index may indicate processes of intense fragmentation of dust component at distances of about 2000 km, or/and an increase in the dust albedo due to the release of the dust from lighter inner parts of the nucleus, or/and evaporation of some darker volatile component of the dust particles in the immediate vicinity of the nucleus.

*Color*

The observed color profiles showed a strong color gradient across the coma: gas-corrected color index BC–RC (i.e., BC corrected for weak emission component and RC corrected for $NH_2$ contamination) was $\sim 1.43^m$ in the innermost near-nucleus area of the comet Encke, and then decreased sharply, reaching a value of about $0.3^m$–$0.4^m$ at the distance of about 2500 km. Dust emitted from the nucleus had a color index almost the same as the cometary nucleus itself, about $1.39^m$. Comparison of the dust color with the color-index of the Sun, which is BC–RC = $1.276^m$ (Farnham et al., 2000), shows that the intrinsic color of the dust changed from slightly red in the vicinity of the nucleus to very blue (color-index about $-1.0^m$) farther from the nucleus. These features indicate some evolution of the scattering particles, most likely, due to changes in composition (evaporation/decomposition of some red organics (Ishiguro et al., 2007)) or/and a change in particle size distribution (increasing abundance of small, Rayleigh-like particles). A similar systematic decrease of the color to the periphery of the coma was also detected in comet 67P/C-G (Rosenbush et al., 2017), despite the fact that the comets are very different: 67P/C-G is a dust-rich comet, and Encke is a dust-poor one.

## Conclusions

It appears that imaging of comets and detailed surface photometry together with simultaneous spectroscopy can yield very valuable information on the structure of the cometary comae and the physical processes in them. In this paper we present some results of the photometric and spectral observations of comet 2P/Encke carried out at the 6-m BTA telescope of the Special Astrophysical Observatory (Russia) on November 4, 2013 and January 23, 2017. Our results can be summarized as follows:

1. In November 2013, comet Encke showed an extended coma with a plasma tail and two jet-like structures. In contrast to this, in January 2017, the images of the comet displayed a strong asymmetric sunward-oriented fan, extending to ~30,000 km from the nucleus. On both dates, there was no regular antisolar dust tail containing small dust particles accelerated by the solar radiation pressure.
2. The radicals CN, $C_3$, $C_2$, and $NH_2$ were the predominant molecular components in the spectrum of comet Encke on January 23, 2017. CH and $CO^+$ molecules were also found in the spectrum. The spectrum showed very little continuum radiation, indicating at low content of dust, and its concentration in the near-nucleus region of the coma.
3. The gas production rates of molecules CN, $C_2$, $C_3$, and $NH_2$ at heliocentric distance 1.052 au were $3.19 \times 10^{25}$ mol/s, $3.26 \times 10^{25}$ mol/s, $0.17 \times 10^{25}$ mol/s, and $0.82 \times 10^{25}$ mol/s, respectively. The values $\log[Q(C_2)/Q(CN)] = -0.08$ and $\log[Q(C_3)/Q(CN)] = -1.00$ for comet Encke corresponded to the typical (not depleted in the carbon-chain) group of comets.
4. The normalized reflectivity gradients are $(9.4 \pm 3.1)$ %/1000 Å and $(8.5 \pm 2.7)$ %/1000 Å from spectroscopy and photometry, respectively. These values are within the range found in the previous determinations, which varied from $(8.9 \pm 1.6)$ to $(13 \pm 7)$ %/1000 Å (A'Hearn et al., 1995).
5. The low dust production was found, although it was somewhat larger than in the previous apparitions of comet Encke. In January 2017, $Af\rho$ determined from spectral data was about $92 \pm 26$ cm and from photometric image in the RC′ filter (i.e. RC corrected for $NH_2$ contamination) was $57 \pm 12$ cm. The dust was concentrated near the nucleus, where the dust/gas ratio was about 3 in the r-sdss filter, however, this ratio continued to be larger than 1 at the distances 3000–40,000 km from the nucleus.
6. We found that about 75% of the flux of the reflected light in the central pixel is due to the nucleus; the nucleus flux contributed 48% in the total intensity of the 2000 km area of the coma. Separating the contributions of the nucleus and the near-nucleus coma, we determined the nucleus magnitude of comet Encke equal to $18.8^m \pm 0.2^m$ that is in a good agreement with the value obtained by Fernández et al. (2000) at the HST for the same observational geometry.
7. Observed color profiles showed a strong color gradient across the coma: color index BC–RC, corrected for the emission component and nucleus contribution, was $\sim 1.43^m$ in the innermost near-nucleus area of the comet, and then decreased sharply, reaching a value of about $0.4^m$ at a distance $\sim 2500$ km. Dust emitted from the nucleus had a color index which was almost the same as the cometary nucleus itself, equal to about $1.39^m$. These features indicate some evolution of scattering particles, most likely, due to changes in their composition or/and particle size distribution.
8. The proposed geometrical model of the dust grains dynamics showed that the observed jets/fan in both observational periods were formed by a single active source located in the north hemisphere at the cometocentric latitude $+55°$.

## Declaration of competing interest

The author declares no conflict of interest.

## Acknowledgments

We thank the anonymous referees for very careful reading of the manuscript and their helpful comments. We appreciate help by Ludmilla Kolokolova in the paper preparation. The observations at the 6-m BTA telescope were performed with the financial support of the Ministry of Education and Science of the Russian Federation (agreement No. 14.619.21.0004, project ID RFMEFI61914X0004). The authors express appreciation to the Large Telescope Program Committee of the RAS for the possibility of implementing the program of observations of comet 2P/Encke at the BTA. The researches by VR, OI, and VK are supported, in part, by the projects 16BF023-02 and 19BF023-02 of the Taras Shevchenko National University of Kyiv. Researches by NK and DP were funded by Russian Foundation for Basic Research grant N⁰ 18-42-910019\18. OI thanks the SASPRO Programme, REA grant agreement No. 609427, and the Slovak Academy of Sciences (grant Vega 2/0023/18). The researches by VR, OI, NK, and OSh were supported, in part, by the Ukrainian-Slovak joint research project for the period 2017–2019.